\documentclass[aps,pre,twocolumn,showpacs,superscriptaddress,final,floatfix]{revtex4-1}
\usepackage{graphicx}  
\usepackage[update]{epstopdf}
\usepackage[inline]{enumitem}

\usepackage{amssymb}
\usepackage{amsmath}
\usepackage{color}
\usepackage{textcomp}
\usepackage[utf8]{inputenc}

\graphicspath {{./figures/}}
\makeatletter
\def\input@path{{./figures/}}
\makeatother

\begin{document}

\title{Approximate ground states of the random-field Potts model from graph cuts}

\author{Manoj Kumar}
\affiliation{School of Physical Sciences, Jawaharlal Nehru University, New Delhi -- 110067, India}
\affiliation{Department of Physics, Indian Institute of Technology, Hauz Khas, New Delhi -- 110016, India}
\author{Ravinder Kumar}
\affiliation{Applied Mathematics Research Centre, Coventry University, Coventry CV1
  5FB, England}
\affiliation{Institut f\"{u}r Theoretische Physik, Leipzig University, Postfach 100920, D-04009 Leipzig, 
  Germany}
\affiliation{Doctoral College for the Statistical Physics of Complex Systems,
  Leipzig-Lorraine-Lviv-Coventry $({\mathbb L}^4)$}
\author{Martin Weigel}
\email{martin.weigel@complexity-coventry.org}
\affiliation{Applied Mathematics Research Centre, Coventry University, Coventry CV1
  5FB, England}
\author{Varsha Banerjee}
\affiliation{Department of Physics, Indian Institute of Technology, Hauz Khas, New
  Delhi -- 110016, India}
\author{Wolfhard Janke}
\affiliation{Institut f\"{u}r Theoretische Physik, Leipzig University, Postfach 100920, D-04009 Leipzig, 
  Germany}
\author{Sanjay Puri}
\affiliation{School of Physical Sciences, Jawaharlal Nehru University, New Delhi -- 110067, India}
\date{\today}

\begin{abstract}

  While the ground-state problem for the random-field Ising model is polynomial, and
  can be solved using a number of well-known algorithms for maximum flow or graph
  cut, the analogue random-field Potts model corresponds to a multi-terminal flow
  problem that is known to be NP hard. Hence an efficient exact algorithm is very
  unlikely to exist. As we show here, it is nevertheless possible to use an embedding
  of binary degrees of freedom into the Potts spins in combination with graph-cut
  methods to solve the corresponding ground-state problem approximately in polynomial
  time. We benchmark this heuristic algorithm using a set of quasi-exact ground
  states found for small systems from long parallel tempering runs. For not too large
  number $q$ of Potts states, the method based on graph cuts finds the same solutions
  in a fraction of the time. We employ the new technique to analyze the breakup
  length of the random-field Potts model in two dimensions.

\end{abstract}

\maketitle

\section{Introduction}
\label{s1}

Due to its versatility, the Potts model is one of the central tools in statistical
physics, in particular for the study of phase transitions and critical phenomena
\cite{wu:82a,potts}. Its many physical realizations include soap froths, cellular
tissues, grain growth, nucleation, as well as static and dynamic recrystallization.
Disorder is inherent in such experimental systems and needs to be incorporated for
their accurate description. Depending on the way the disorder couples to the system,
this leads to the $q$-state random-bond or random-field Potts models. Experimentally,
the latter is particularly relevant for describing magnetic grains, anisotropic
orientational glasses, randomly diluted molecular crystals~\cite{br92,michel},
structural transitions in SrTiO$_{3}$ crystals~\cite{aharony}, and phase transitions
in type I antiferromagnets (such as NdSb, NdAs, CeAs) in a uniform
field~\cite{dsm}. While the random-bond model has received substantial attention in
the past and is relatively well understood in two (2d)
\cite{chen:92,picco:97,berche:03a} as well as in three (3d) dimensions
\cite{chatelain:01a,berche:02a,hellmund:03a,chatelain:05}, little is known about the
behavior of the random-field Potts model (RFPM). Due to the necessary quenched
average over disorder and the slow relaxation resulting from the frustration
introduced through the competition of exchange couplings and random fields (of
strength $\Delta$), it is a difficult problem for analytical and numerical methods
alike.  Consequently, the nature of phases and the phase transitions in the
$(q,\Delta)$-plane in different spatial dimensions $d$ are only partially understood,
leaving many open questions for exploration.

There are only a few studies of the $q$-state RFPM in the
literature~\cite{blankschtein:84,nishimori,eichhorn:95,eichhorn:95a,eichhorn:96,reed}. These
have primarily investigated the phase diagram in $(T,\Delta)$-space for different $q$
and $d$, where $T$ denotes temperature. For the pure model, the temperature-driven
transitions are continuous for small $q \le q_c$ and first-order for large $q > q_c$,
with $q_c = 4$ for the square-lattice model with nearest-neighbor interactions
\cite{duminil:15b,duminil:16} and $q_c \approx 2.8$ for the simple-cubic lattice
\cite{hartmann:05}. It is well known that quenched disorder tends to soften
first-order transitions \cite{cardy:99a}, and this has even been rigorously
established for systems in two dimensions \cite{aizenmann:89a}. In the latter case,
one hence must have $q_c\to\infty$, but in fact the RFPM does not show a
finite-temperature ordering transition in $d=2$ and there are merely crossovers
between ferromagnetic and paramagnetic states for finite systems, much alike to the
behavior found for the 2d random-field Ising model (RFIM) \cite{binder:83}. For
$d>2$, on the other hand, one would at least expect for $q_c$ to increase on coupling
to the disorder. The only numerical studies of the problem are due to Eichhorn and
Binder \cite{eichhorn:95,eichhorn:95a,eichhorn:96}, who considered the case $q=3$ and
$d=3$ using Monte Carlo simulations. They proposed a qualitative scenario in the
$(q,d)$-plane which exhibited a shift of the tri-critical curves $q_c(d)$ to higher
values, consistent with the mean-field predictions of Blankschtein {\em et al.}
\cite{blankschtein:84}. The simulation results for $q=3$ indicated a continuous
transition for the considered disorder strength. It is clear, however, that these
simulations, which date back to before the advent of modern simulation techniques for
disordered systems such as parallel tempering \cite{hukushima:96a}, might be affected
by equilibration problems and strong corrections to finite-size
scaling. Additionally, the question of whether a softening of discontinuous
transitions occurs for all strengths $\Delta$ of the random fields or only above a
certain threshold has not been addressed to date.

A related system is the random-field Ising model (RFIM) \cite{nattermann:97} which,
up to a rescaling, can be mapped onto the RFPM for $q=2$, see the discussion in
Sec.~\ref{s2}. Although this system was studied extensively over the past decades, it
is only recently that large-scale numerical studies were able to settle a number of
important questions for this problem \cite{fytas:13,fytas:16}, such as the number
(and values) of independent exponents, the universality of transitions with respect
to the coupling distribution and the issue of dimensional reduction
\cite{parisi:79a}. An important feature of the renormalization-group (RG) treatment
of this system is that the RG fixed point that controls the disordered transition is
located at $T=0$ \cite{bray:85a}. As a consequence, a systematic study of ground
states in this case allows to extract the critical exponents of the transition at
finite temperature so it exists. It is a fortunate coincidence that the problem of
finding the ground state for an RFIM sample can be mapped to a maximum-flow problem
that is in P \cite{dauriac:85}, i.e., there exist algorithms that solve it in a time
that grows as a polynomial in the size of the system, including the Ford-Fulkerson
algorithm of augmenting paths~\cite{ff5}, the Goldberg-Tarjan push-relabel
method~\cite{gt5}, or variants thereof \cite{kolmogorov:04}. In the last few years,
we have acquired significant knowledge about the ground-state properties of the
RFIM~\cite{epl96,pre90,epj37,ijp88,stevenson:11,stevenson:11a}. The situation is
different, however, for the case of the RFPM with $q> 2$ which corresponds to a
multi-terminal flow or, equivalently, graph cut (GC) problem that is known to be NP
hard \cite{bvz,dauriac:85}. Still, as was shown by Boykov {\em et al.\/} \cite{bvz},
solutions to such multi-terminal flow problems can be efficiently approximated using
an embedding of binary degrees of freedom into the states with more than two labels.

In the present paper, we undertake a first exploratory study into determining ground
states of the $q$-state RFPM using graph-cut methods. By comparing the results of the
heuristic GC algorithm to those of parallel tempering (PT) simulations systematically
tuned to yield ground states for small systems in 2d with very high success
probabilities, we establish that the GC approach yields reasonable estimates of
ground states for the $q$-state RFPM. The run times of the GC approach are
significantly smaller than those of the PT simulations, and they scale linearly with
the system size as well as the number of states, allowing us to study large system
sizes.

The rest of the paper is organized as follows. In Sec.~\ref{s2}, we describe the
$q$-state random field Potts model, the graph cut method, and the parallel tempering
approach. Section~\ref{s3} provides detailed comparisons between (quasi) {\em
  exact\/} ground states obtained using PT and {\em approximate\/} ground states
found using the GC method. We also demonstrate here that GC provides a good
approximation to the ground states, especially for small $q$.  In
Sec.~\ref{sec:breakup} we apply the GC method to study the breakup length for the
$q=3$ and $q=4$ RFPM in two dimensions. Finally, in Sec.~\ref{s4}, we conclude this
paper with a summary and discussion.

\section{Model and Methodology}
\label{s2}

In the following, we describe the variant of the RFPM studied here and introduce two
numerical approaches for determining ground states of samples, the graph-cut method
and parallel tempering.

\subsection{Random Field Potts Model}

The ferromagnetic $q$-states Potts model is described by the Hamiltonian
\cite{wu:82a}
\begin{equation}
  {\cal H} = -J\sum_{\langle ij\rangle} \delta_{s_i,s_j},
\end{equation}
where the $s_i\in \{0,1,\ldots,q-1\}$ are the Potts spins, $\langle ij\rangle$
denotes summation over nearest neighbors only, and $J > 0$ is a (ferromagnetic)
coupling constant. For the purposes of the present study, we consider systems on
square and simple-cubic lattices with periodic boundary conditions. The coupling of
the spins to random fields can take a variety of different forms
\cite{blankschtein:84,eichhorn:96,gx}. A symmetric coupling of continuous fields can
be expressed as \cite{blankschtein:84}:
\begin{equation}
  \label{hamilt}
  \mathcal{H}=-J\sum_{\left<ij\right>}\delta_{s_i,s_j}-
  \sum_i\sum_{\alpha=0}^{q-1}h_{i}^{\alpha}\delta_{s_i,\alpha},
\end{equation}
where $\{h_{i}^{\alpha}\}$ denotes the quenched random field at site $i$, acting on
state $\alpha$. Hence, in this model, the random field at each site has $q$
components, and we take each of these to follow a normal distribution. To separate
the disorder strength from the random instance we define
$h_{i}^{\alpha} = \Delta\epsilon_i^\alpha$ and $\epsilon_i^\alpha$ are then drawn
from a standard normal distribution, i.e.,
\begin{equation}
  P(\epsilon_i^\alpha) = \frac{1}{\sqrt{2\pi}} \exp \left(-{\epsilon_i^\alpha}^2/2 \right).
  \label{eq:ditrib}
\end{equation}
For the case $q=2$, the Hamiltonian (\ref{hamilt}) has two different random fields
$h_i^0\equiv h_i^+$ and $h_i^1 \equiv h_i^-$ for the two spin orientations, in
contrast to the usual definition of the RFIM \cite{nattermann:97}. As is easily seen,
however, Eq.~(\ref{hamilt}) in this case can be written as
\begin{equation}
  \mathcal{H}=-\frac{J}{2}\sum_{\left<ij\right>}[\sigma_i\sigma_j+1]
  -\frac{1}{2}\sum_i[(h_i^+-h_i^-)\sigma_i+(h_i^++h_i^-)],
\end{equation}
where $\sigma_i=\pm 1$ are Ising spins. It is hence clear that, up to a constant
shift, the $q=2$ RFPM of Eq.~(\ref{hamilt}) and with the distribution
(\ref{eq:ditrib}) at coupling constant $J$ and random field $\Delta$ is equivalent to
the RFIM at coupling $J/2$ and field strength $\Delta/\sqrt{2}$.

An alternative model with discrete distribution of the disorder is given by
\cite{eichhorn:96,gx}
\begin{equation}
  \label{hamilt_dis}
  \mathcal{H}=-J\sum_{\left< ij\right>}\delta_{s_i,s_j}-\Delta \sum_i\delta_{s_i,h_i}.
\end{equation}
Here, the quenched random variables $h_i$ are chosen uniformly from the set
$\{0,1,\ldots,q-1\}$, i.e.,
\begin{equation}
  P(h_i)=\frac{1}{q}\sum_{\alpha=0}^{q-1} \delta_{h_i,\alpha}.
\end{equation}
Thus the distribution of random fields is discrete, and couples to any one of the $q$
spin states with equal probability. We note that for the continuous form
(\ref{hamilt}) we expect a unique ground state, while the alternative
(\ref{hamilt_dis}) might admit (extensive) degeneracies, in particular for rational
choice of $\Delta$. While the discreteness of the form (\ref{hamilt_dis}) might have certain advantages
for the efficient implementation of simulation codes, we would like to avoid the
possible subtleties associated with degeneracies, and we will hence use the form
(\ref{hamilt}) here.

\subsection{Graph Cut Method}
\label{sec:gc}

While the ground-state problem of the RFPM is NP hard and so an exact solution is out
of reach, an efficient approximation algorithm has been developed in the context of
applications of the Potts model (and related systems) in computer vision
\cite{bvz}. The method is tailored for a general energy function of the form
\begin{equation}
  E(\{s_{i}\}) =  \sum_{i,j} V_{ij}(s_i,s_j) + \sum_{i}D_i(s_i), 
  \label{eq:veksler}
\end{equation}
where in the original application $s_i$ would have referred to the color label of the
pixels of a (planar) image, but the interaction matrix $V_{ij}$ can be such that also
more general graphs and three-dimensional systems can be modeled. The RFPM
Hamiltonian (\ref{hamilt}) is clearly a special case of this general form. Each site
$i$ is assigned a label $s_i \in \{0,1,\ldots,q-1\}$. The function $V_{ij}(s_i,s_j)$
gives the cost of assigning labels $s_i$ and $s_j$ to the sites $i$ and $j$, while
the function $D_i$ measures the penalty (or cost) of assigning the label $s_i$ to
site $i$. The basic approach taken in Ref.~\cite{bvz} is to consider constraint
optimization problems derived from Eq.~(\ref{eq:veksler}) in such a way that the $q$
labels are reduced to an effective two-label problem. As this is then equivalent to
(a slight generalization of) the RFIM, a ground state for this constraint problem can
be determined exactly and in polynomial time using the established min-cut/max-flow
algorithms. This idea is in the same spirit as the embedding of Ising variables used
in combination with minimum-weight perfect matching in dealing with continuous-spin
glasses on planar lattices \cite{weigel:05f,weigel:06b}.

The two approaches of this type proposed in Ref.~\cite{bvz} are the
$\alpha$-$\beta$-swap and the $\alpha$-expansion moves. For the $\alpha$-$\beta$-swap
one picks two labels $\alpha \ne \beta \in \{0,1,\ldots,q-1\}$ and freezes all labels
apart from $\alpha$ and $\beta$. Under this constraint, the problem
(\ref{eq:veksler}) is equivalent to a two-label problem on the sites with labels
$\alpha$ or $\beta$ that can be solved by min-cut/max-flow. This step is repeated for
each pair of labels, resulting in a {\em cycle\/} of $q(q-1) \approx q^2$ steps. For
the $\alpha$-expansion move one picks a label $\alpha$ which is then frozen. The
remaining pixels are given the alternative of either keeping their current label or
being flipped into the $\alpha$ state, which is again a binary choice, and the
resulting constraint problem can be solved by min-cut/max-flow. A cycle of the
$\alpha$-expansion takes $q$ steps. Independent of which of the two algorithms is
used, cycles are repeated until the configurations do not change any further, and the
methods have converged to a local minimum.

While these algorithms are not exact and are hence not guaranteed to find ground
states, they have been reported to yield excellent approximations to the ground
states and are widely used in computer vision. For the $\alpha$-expansion move, it is
possible to derive an upper bound on the energy of the local minima found, which is
given by
\begin{equation}
  E(\hat{f})\leq 2cE(f^*), \quad  \mbox{where} \quad c=\frac{{\rm max}_{s_i\neq s_j}
    V(s_i,s_j)}{{\rm min}_{s_i\neq s_j}V(s_i,s_j)},
\end{equation}
$\hat{f}$ is the state returned by the $\alpha$-expansion move, and $f^*$ is the
global optimum.  For the Potts model, $V_{ij}(s_i,s_j)\equiv -J\delta_{s_i,s_j}$,
yielding $c=1$. So the expansion move provides a local minimum within a factor of two
of the global minimum. To check the effectiveness of their algorithms, the authors of
Ref.~\cite{bvz} experimented on a variety of computer vision problems such as image
restoration with multiple labels, stereo and motion. These problems are solved by
computing a minimum cost {\it multi-way cut} on the graph. A comparison of their
results with known ground states revealed 98\% accuracy \cite{bvz}. The method has
not previously been applied to the RFPM, and to benchmark it there, we need a
collection of samples with known ground states. For this purpose, we use the replica
exchange or {\it parallel tempering} (PT) method, which can be used to find exact
ground states with high probability for small systems.

\subsection{Parallel Tempering}
\label{sec:tempering}

\begin{table}[tb!]
  \caption{Optimized values of $\eta$ according to Eq.~(\ref{Tm}) and
    the number of temperature replicas $N_T$ for
    different lattices with $L^2$ spins used in the parallel tempering.}
  \label{table:parameter}
  \begin{center}
    \begin{tabular}{ c cccccc c}
      \hline\hline
      $L$ & 8 & 12 & 16 &20 &24 &32 &40\\ 
      \hline
      $N_T$ & 16& 16 & 16 &16 &16 &32 &32 \\ 
      $\eta$ & 1.13& 1.13 & 1.13 &1.13 &1.13 &1.14 &1.14 \\ 
      \hline\hline
    \end{tabular}
  \end{center}
\end{table}

Ground states of RFPM samples could be generated via exact enumeration of states. Due
to their exponential number $\sim q^N$ this only works for the tiniest of systems,
however. While this situation could possibly be improved with the use of
branch-and-cut techniques \cite{simone:95}, we do not follow this approach here and
instead revert to stochastic approximation schemes based on Markov chain Monte Carlo
simulations. Simple Monte Carlo at a fixed low or even zero temperature will not lead
to ground states.  The RFPM has a complicated free energy landscape with many minima
and maxima. These metastable states trap the evolving system and impede the
relaxation to the ground state. Any reasonable Monte Carlo sampling therefore has to
overcome energy barriers and cross from one basin to another to reach the global
minimum. Established approaches to achieve this are simulated annealing
\cite{kirkpatrick:83} and parallel tempering \cite{geyer:91,hukushima:96a}. It has
been shown that among the Monte Carlo methods parallel tempering consistently
outperforms simulated annealing as a tool for ground-state searches in disordered
systems \cite{wang:15} \footnote{But note that population annealing
  \cite{hukushima:03,machta:10a} might be another interesting contender in this
  respect \cite{wang:15,barash:16}.}.  We will hence focus on parallel tempering
(PT).

Consider $N_T$ initially non-interacting replicas of the system at distinct
temperatures. In PT each replica is evolved at its temperature $T_m$ using canonical
Monte Carlo, for example employing the single spin-flip Metropolis method
\cite{binder:book2}. In an additional step, replicas with neighboring temperatures
are exchanged with the probability
\begin{equation}
  P_{\rm ex}=\min\left[1, e^{(\beta_{m}-\beta_{m+1})(E_m-E_{m+1})}\right],
  \label{ptaccept}
\end{equation}
where $\beta_m=1/k_B T_m$ and $E_m$ denotes the configurational energy of the $m$th
replica. This scheme couples the replicas and allows copies that are trapped in
metastable states at low temperatures to escape to high temperatures via successive
exchanges with neighboring copies, where they can more easily relax to then return
via the same random walk in temperature space to low temperatures, typically
exploring a different basin.

\begin{figure}[tb!]
  \centering
  \includegraphics[width=0.6\columnwidth]{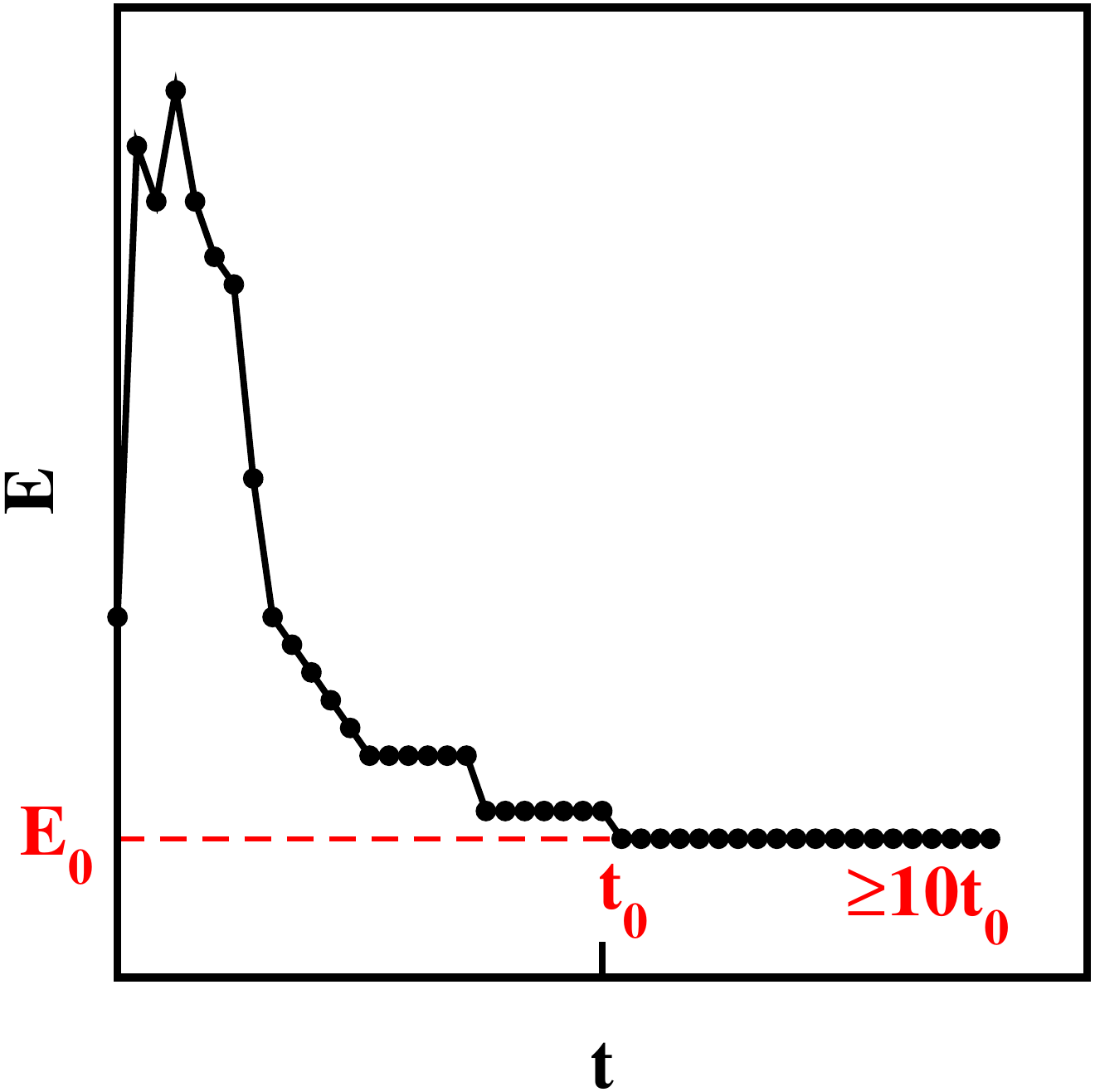}
  \caption{Schematic diagram showing the variation of $E$ with time $t$ in the PT
    runs (in MC steps). The first occurrence of the minimum energy $E_0$
    is at onset time $t=t_0$. The corresponding state is accepted as a ground state if no
    lower energy is found up to $t\ge 10 t_0$.}
  \label{demo_t0}
\end{figure}

The most delicate aspect of PT relates to the choice of the number and spacing of the
replicas in temperature space. Clearly, neighboring temperatures must be close enough
such that the acceptance probabilities (\ref{ptaccept}) are appreciable, which
essentially means that the energy histograms at neighboring temperatures must have
sufficient overlap \cite{bittner:08}. On the other hand, too many replicas with
consequently high acceptance rates of swaps are also not ideal as this slows down the
random walk in temperature space through smaller and smaller temperature steps. A
number of different protocols have been suggested for choosing the optimal set
$\{T_m\}$
\cite{katzgraber:06,bittner:08,kofk02,kon05,pred04,sabo08,neiro00,calvo05,brenner07}. Here
we use a heuristic scheme based on a generalization of the widely used geometric
progression of temperatures \cite{katzgraber:09a}, and choose the temperatures
according to
\begin{equation}
  \label{Tm}
  T_m=m^{\eta} T_{\rm norm} + T_{\rm min}, \quad \mbox{where}
  \quad T_{\rm norm}=\frac{T_{\rm max}-T_{\rm min}}{(N_T-1)^{\eta}}.
\end{equation}
The choice of the maximum and minimum temperatures $T_{\rm max}$ and $T_{\rm min}$ is
guided by the need to select a sufficiently high $T_{\rm max}$ to ensure good
relaxation of replicas that arrive there, and (in our case of using PT as a global
optimization algorithm) a sufficiently low $T_{\rm min}$ to allow us to find ground
states. From preliminary tests, we found that $T_{\rm max} =1.5$ and
$T_{\rm min} =0.2$ are sufficient for our purposes. We first determine the number of
replicas $N_T$ by generating the corresponding sequence of temperatures for
$\eta=1$. If a test run shows overall low swap acceptance rates, we increase
$N_T$. The adjustable parameter $\eta$ is found recursively as follows: (1) The
simulations are performed for a chosen value of $\eta$ and the set of temperatures
$\{T_m\}$ determined using Eq.~(\ref{Tm}); (2) The tunneling time, i.e., the average
time for a replica to travel from the lowest to the highest temperature and back, is
measured in a test simulation \footnote{In a somewhat simpler setup, one could also
  just monitor the swap acceptance rates.}; (3) Steps 1 and 2 are repeated for a
modified value of $\eta$. The value of $\eta$ which minimizes the tunneling time is
selected to yield the optimal set $\{T_m\}$. The values of $\eta$ and $N_T$ for a
lattice of lateral size $L$, after this optimization protocol, are listed in
Table~\ref{table:parameter} \footnote{Note that we did not fully follow the
  optimization procedure described above, but only considered a few discrete choices
  of $\eta$.}. For simplicity we used the same parameters for different numbers of
states $q$, although for best performance these cases should be separately optimized.

\begin{figure}
  \centering
  \includegraphics[width=0.95\columnwidth]{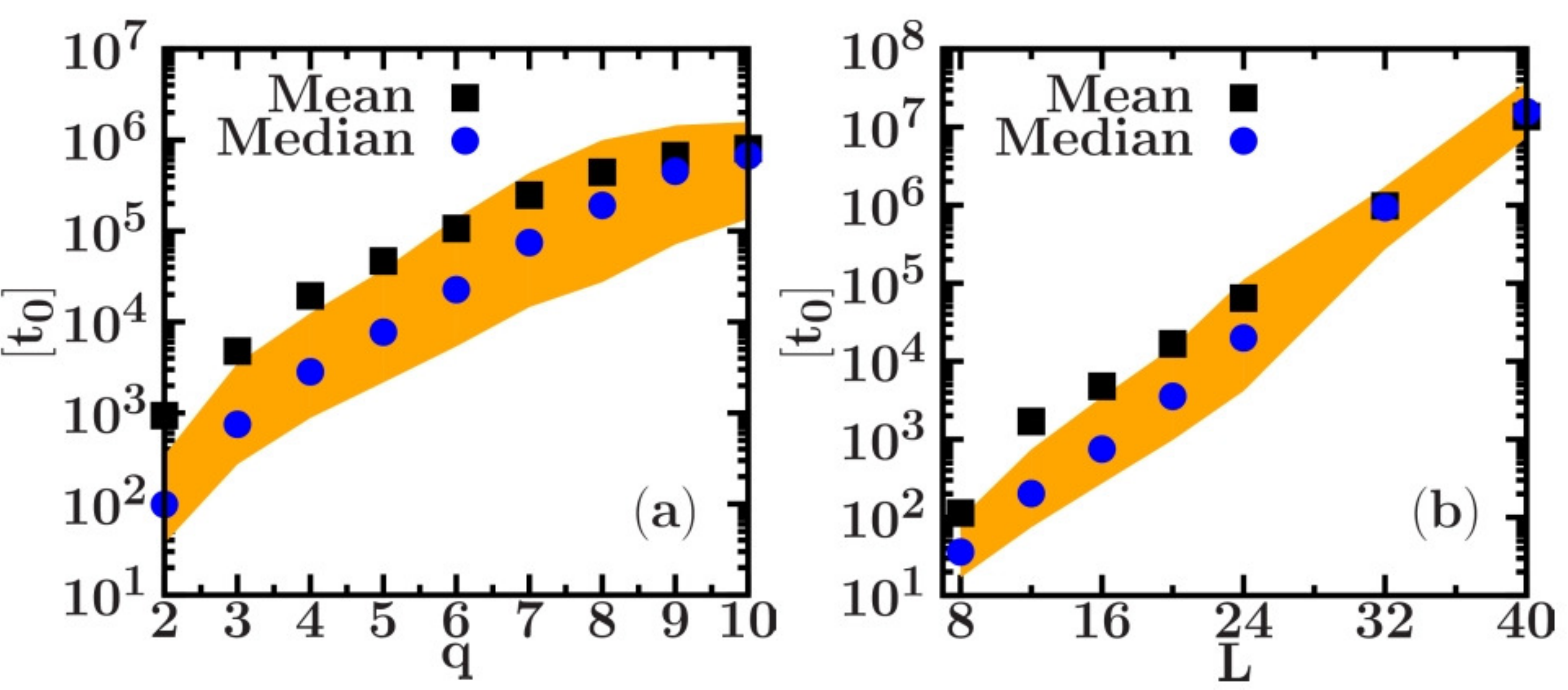}
  \caption{Disorder-averaged onset times $t_0$ for finding the ground states of the
    $d=2$ RFPM using parallel tempering.  The data is plotted on a linear-log scale
    as a function of (a) the number of Potts states $q$ for $L=16$, and (b) the
    system size $L$ for $q=3$. The data are averaged over $1536$ realizations of
    quenched random fields according to Eq.~(\ref{eq:ditrib}) with $\Delta = 1$. The
    shaded area shows the range that contains the onset times for 66\% of the samples.
  }
  \label{fig:first_hit}
\end{figure}

\begin{figure*}[tb!]
  \centering
  \includegraphics[width=0.75\linewidth]{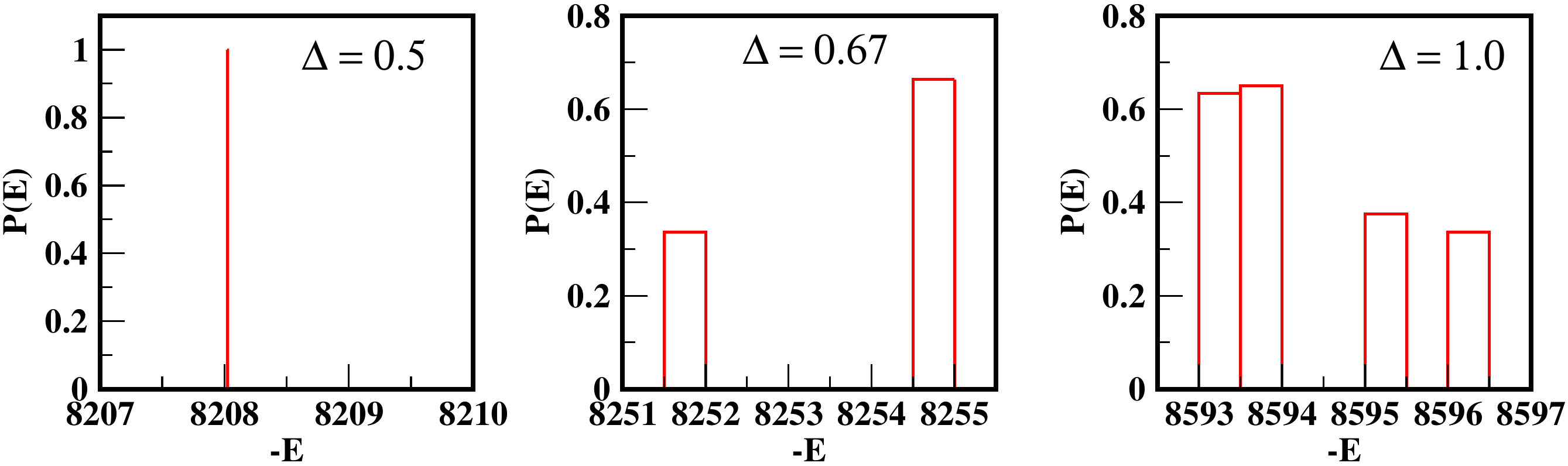}
  \includegraphics[width=0.75\linewidth]{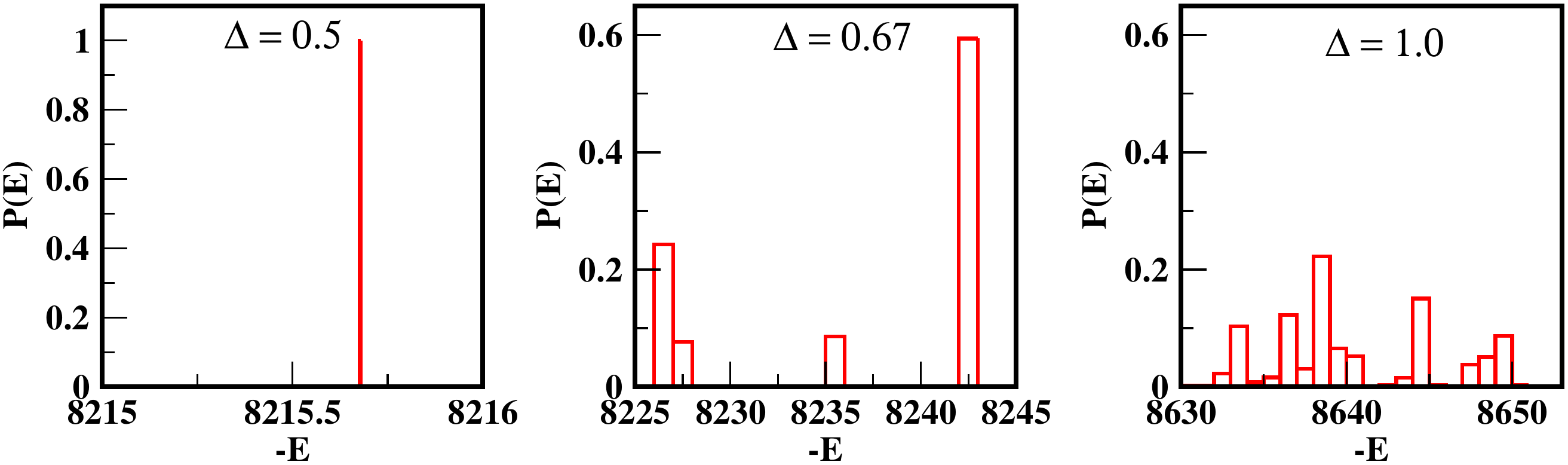}
  \caption{Energy histograms of final states obtained from GC for the $q=3$ RFPM (top
    row) and the $q=4$ RFPM (bottom row) on a $64^2$ lattice. The histograms are
    obtained from 10\,000 initial configurations $\{s_i\}$ for a fixed disorder
    configuration $\{\epsilon_i^\alpha\}$.}
  \label{hist_q34}
\end{figure*}

As a realization of Markov chain Monte Carlo that satisfies ergodicity and detailed
balance, PT is guaranteed to converge to the equilibrium distribution
\cite{binder:book2}. Nevertheless, while it performs much better than local updates
alone, for systems with complex free-energy landscapes such as the RFPM the
equilibration times can still be very long, and they increase steeply with system
size and with lowering $T_\mathrm{min}$. For not too large systems, however, we are
able to find ground states for the overwhelming majority of samples. To ensure this,
we rely on the following bootstrapping procedure:
\begin{enumerate}
\item We run all samples for given $L$ and $q$ for some initial time chosen to ensure
  equilibration of an average sample (determined, for example, by measuring the
  average tunneling time).
\item For each sample, we determine the onset time $t_0=t_0(\{h_i^\alpha\})$, i.e.,
  the time when the lowest energy seen in the whole run is observed first.
\item We re-run each sample with a runtime of
  $t(\{h_i^\alpha\})=10\times t_0(\{h_i^\alpha\})$.
\item For samples where a new, lower state is found in the extended runs, we repeat
  this procedure until the condition  $t=10\times t_0$ is met.
\end{enumerate}
This procedure is illustrated in Fig.~\ref{demo_t0}. It is highly reliable in finding
ground states, and we estimate the failure probability for the system sizes
considered to be of the order of 1 in 1000. For none or the samples considered here
was a state lower than the reference state determined from the procedure above found
in any of the other runs (PT or GC).

We performed such simulations for system sizes $8\le L\le 40$ and number of states
$2\le q\le 10$ for 1536 configurations of the random fields each. The resulting
average and median onset times of the ground states are shown in
Fig.~\ref{fig:first_hit}.  The shaded area indicates the level of disorder
fluctuations.  These plots are shown on a linear-log scale.  In
Fig.~\ref{fig:first_hit}(a), we observe that $t_0$ increases slightly slower than
exponentially with the number of Potts states $q$.  In Fig.~\ref{fig:first_hit}(b),
we observe an exponential increase of $t_0$ for system sizes $L \ge 16$. This is what
we expect for any process geared towards ensuring exact ground states as the problem
is NP hard. As the mean values are larger than the medians, the distribution is
asymmetrical and tail-heavy for all values of $q$ and $L$.

\section{Benchmarks}
\label{s3}

We first consider the behavior of the GC method in its own right before turning to a
detailed comparison of this technique to the PT method. The bulk of our runs were
performed in two dimensions, but some of the timing runs discussed in
Sec.~\ref{sec:timing} were repeated for cubic lattices.

\subsection{Approximate ground states from GC}

\begin{figure}[tb!]
  \centering
  \includegraphics[width=.95\columnwidth]{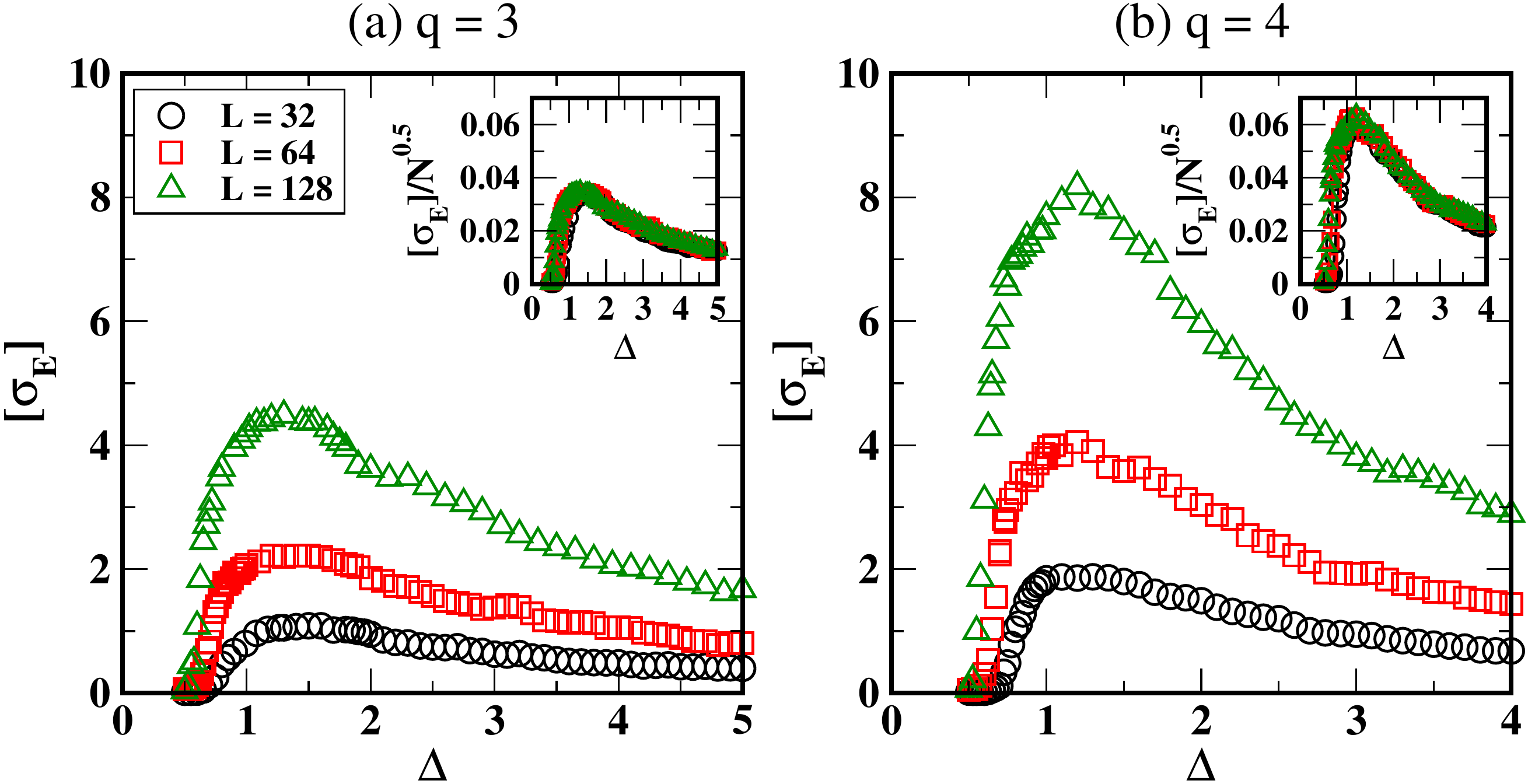}
  \caption{Variation of the standard deviation of energy $[\sigma_E]$ from runs of
    the GC method as a function of disorder amplitude $\Delta$ for the $d=2$ RFPM
    with (a) $q=3$, and (b) $q=4$. All data are averaged over 100 disorder
    realizations $\{\epsilon_i^\alpha\}$, and 1000 initial states $\{s_i\}$ for
    each disorder realization. Clearly, $[\sigma_E]$ grows with lattice size $L$,
    and also with number of states $q$. The scaled data in the insets demonstrate
    that $[\sigma_E] \sim \sqrt{N} = L$, i.e., there are no critical fluctuations
    in this range of $\Delta$-values.}
  \label{ene_sp_2d}
\end{figure}

Let us begin by testing the final states obtained via graph cuts in the RFPM. We fix
the disorder configuration $\{\epsilon_i^\alpha\}$ and obtain the final states from
several runs of GC for different initial spin configurations $\{s_i\}$. The top row
of Fig.~\ref{hist_q34} shows energy histograms of these states for the $q=3$
RFPM. The simulations are performed on a $64^2$ lattice with $10\,000$ initial
conditions.  For $\Delta=0.5$, the system always converges to the same energy
state. Hence the histogram shows a sharp peak corresponding to that energy state. As
we increase the disorder strength $\Delta$, the distribution spreads over multiple
energy states. The bottom row of Fig.~\ref{hist_q34} shows similar histograms for the
$q=4$ RFPM, which are even wider. Therefore, the GC method is not guaranteed to yield
a ground state of the RFPM. This is corroborated by a comparison of the actual
states found to the true ground states as discussed in Sec.~\ref{sec:comparison}
below.

To quantify the energy spread in the histogram, we determine the standard deviation
in energy,
\begin{equation}
  \sigma_E=\left(\left<E^2\right>-\left<E\right>^2\right)^{1/2},
\end{equation}
where the angular brackets $\left<\cdot\right>$ denote an average over different
initial conditions for a fixed disorder realization. A further average over
independent disorder configurations yields the disorder-averaged quantity
$\left[\sigma_E\right]$. In Fig.~\ref{ene_sp_2d}, we plot $\left[\sigma_E\right]$
vs.\ $\Delta$ on a $d=2$ lattice ($L\times L\equiv N$) for $L=32$, $64$, $128$. The
data has been averaged over 100 disorder realizations, and 1000 initial states for
each disorder configuration. The energy spread grows with $L$. To understand this
dependence, we plot $\left[\sigma_E\right]/\sqrt{N}$ vs.\ $\Delta$ in the insets. The
data collapse shows that $\left[\sigma_E\right] \sim \sqrt{N}$, demonstrating the
absence of critical fluctuations. The relative fluctuations in the energy,
$\left[\sigma_E\right]/\left<E\right>\sim N^{-1/2}$, vanish in the thermodynamic
limit. In the limit $\Delta \rightarrow \infty$, we can neglect the exchange term in
Eq.~\eqref{hamilt}, which yields $\left<E\right> = -N \Delta$ and
$\left<E^2\right> = N^2 \Delta^2$, i.e., $\sigma_E \rightarrow 0$ as
$\Delta \rightarrow \infty$.

\begin{figure}[tb!]
  \centering
  \includegraphics[width=.6\columnwidth]{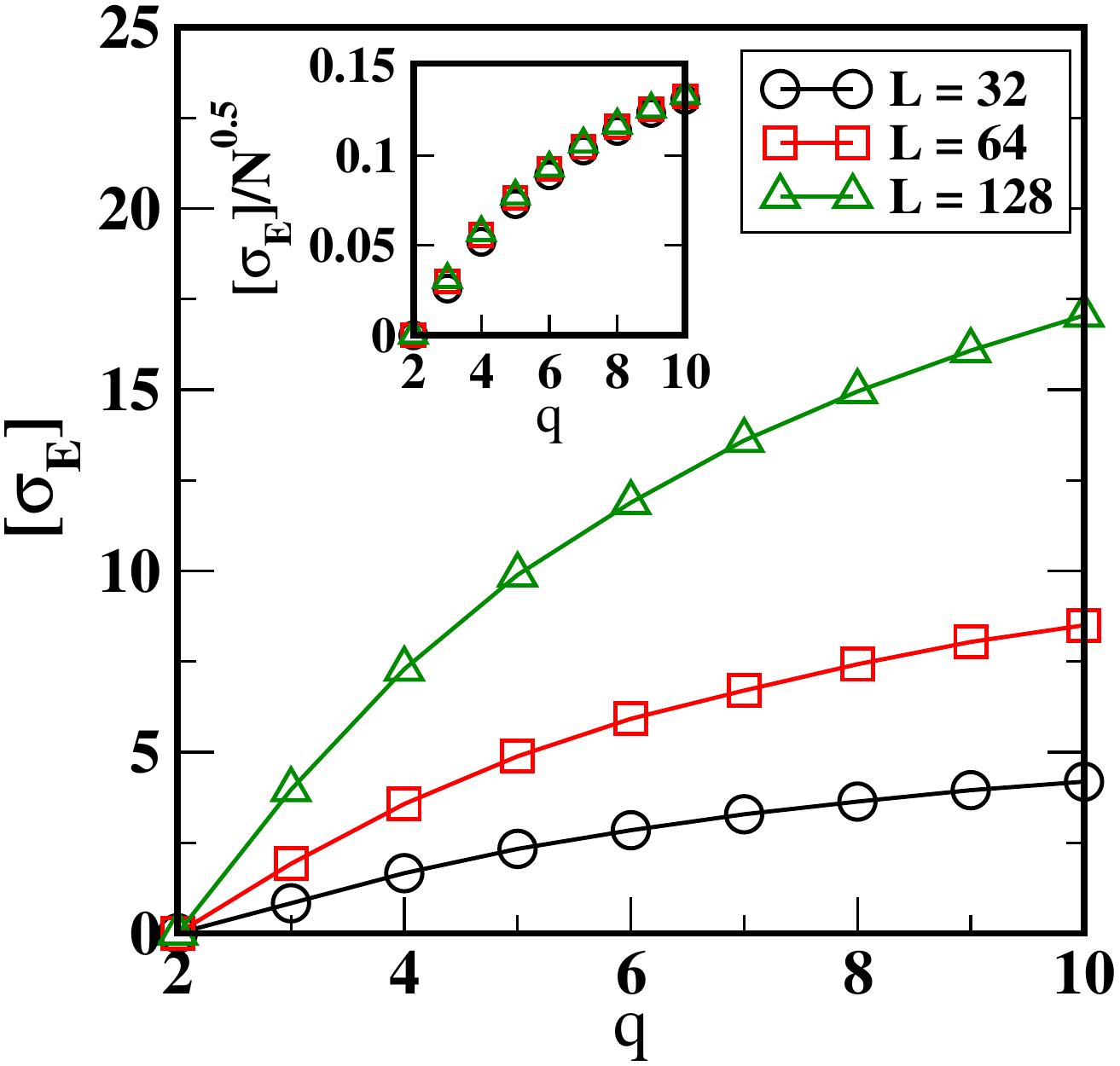}
  \caption{Plot of $[\sigma_E]$ vs.\ $q$ for the $d=2$ RFPM with $\Delta=1.0$, and
    indicated lattice sizes. The statistics is the same as in Fig.~\ref{ene_sp_2d}.
    The spread in energy of the GC states increases with $q$.
    The inset shows that $[\sigma_E] \sim \sqrt{N} = L$.
  }
  \label{ene_sp_q}
\end{figure}

The $q$-dependence of the energy spread can be understood from Fig.~\ref{ene_sp_q},
where we plot $\left[\sigma_E\right]$ vs.\ $q$. The data sets correspond to
$\Delta = 1.0$. The increase in number of metastable states with $q$ implies that the
GC approach becomes worse in terms of the quality of the energy minima. The inset of
this figure again confirms $\sigma_E\sim\sqrt{N}$.

\subsection{Comparisons with PT}
\label{sec:comparison}

\begin{figure}[tb!]
  \centering
  \includegraphics[width=.475\columnwidth]{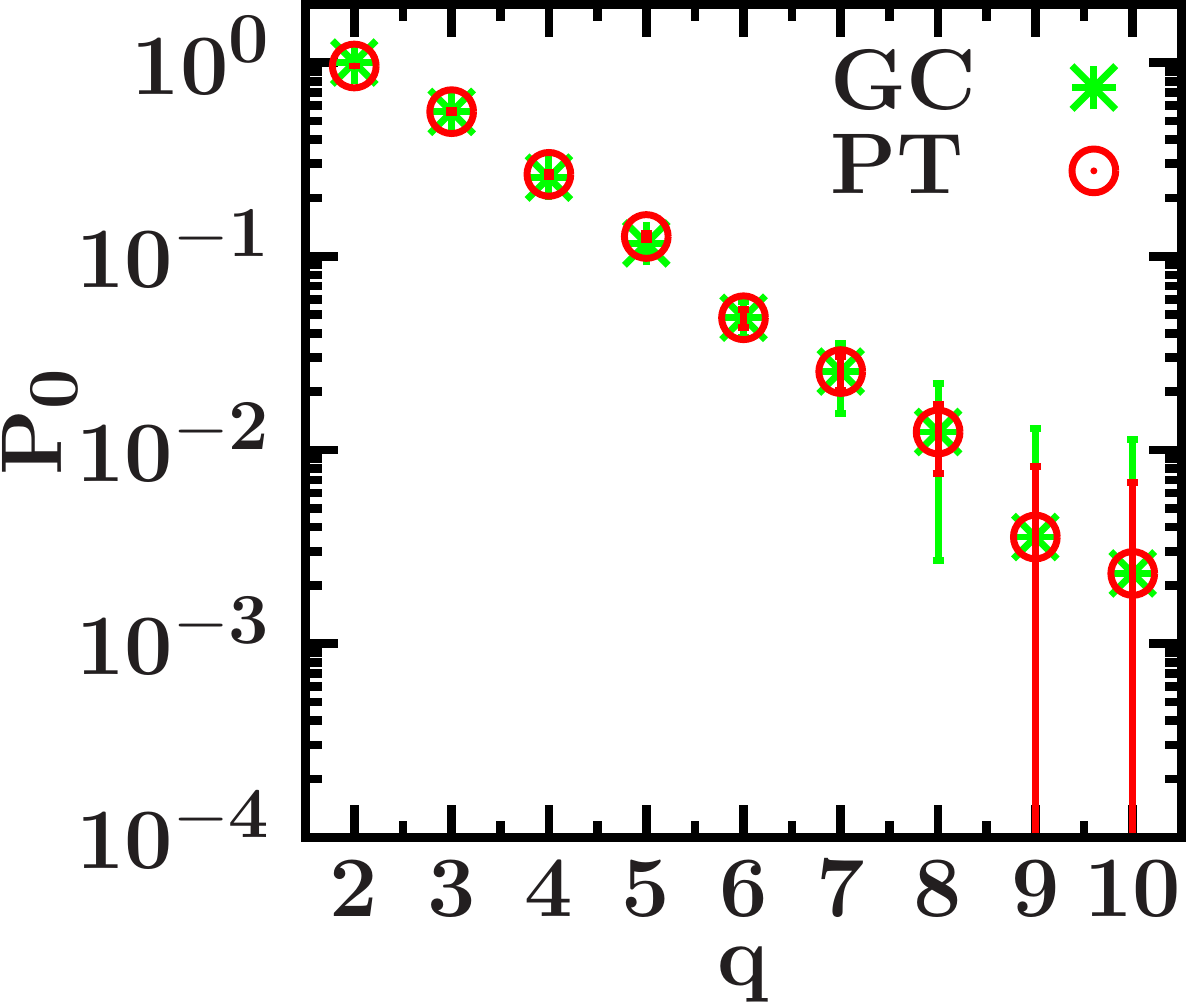}
  \includegraphics[width=.475\columnwidth]{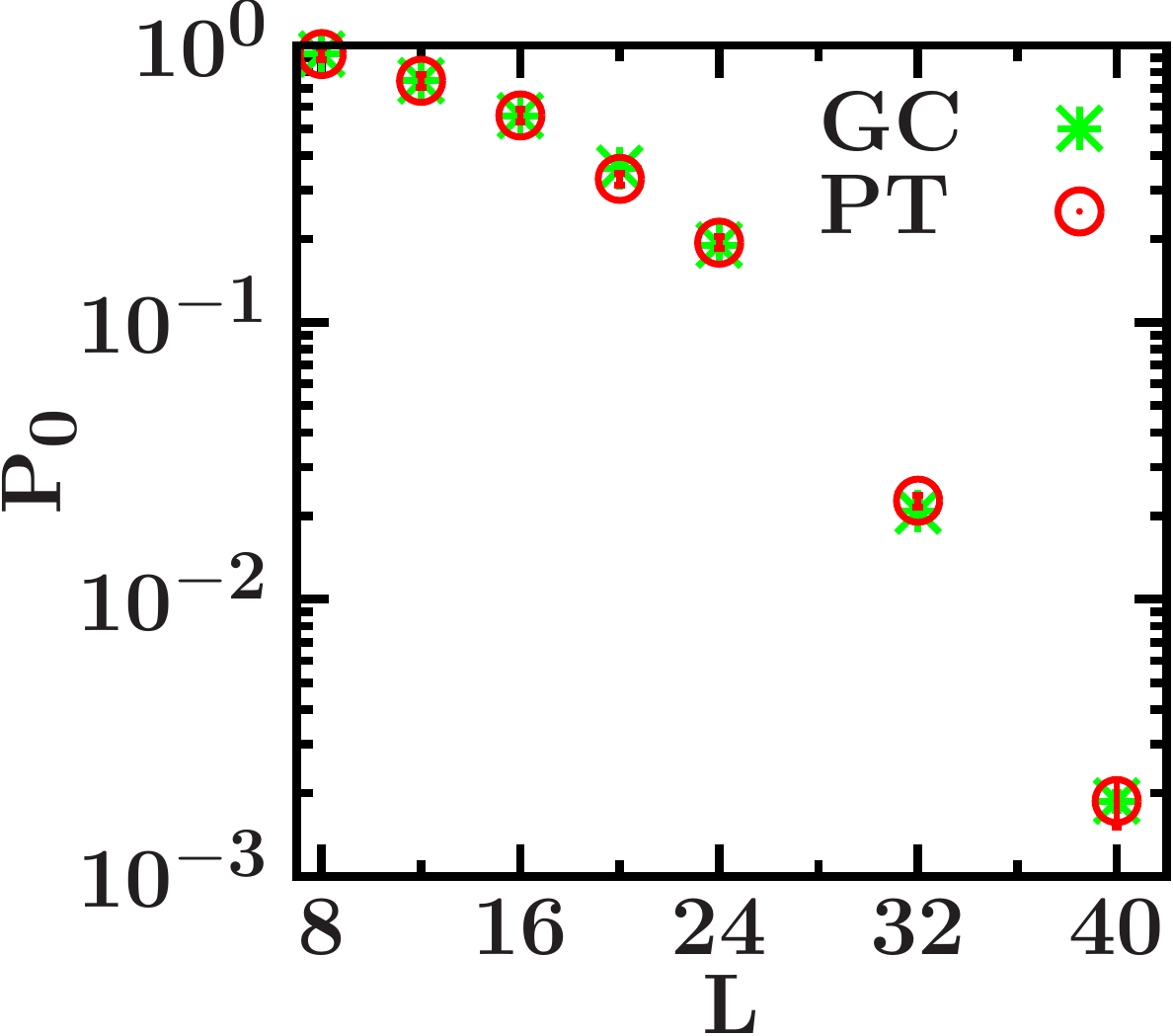}
  \caption{ Disorder-averaged success probability of finding the ground state from GC
    and PT runs for 2d RFPM as a function of $q$ (left panel, $L=16$) and $L$ (right
    panel, $q=3$). The GC data correspond to one initial state per disorder sample,
    while the run time in PT was adapted to yield exactly the same average success
    probability as the corresponding GC run (see main text). All data are averaged
    over 1536 configurations of the random fields.  }
  \label{fig:success}
\end{figure}

Having established a database of samples for which the ground states are known with
very high probability through the PT procedure described in Sec.~\ref{sec:tempering},
it is possible to benchmark the GC method against quasi-exact results as well as
against PT runs. In Fig.~\ref{fig:success} we show the average success probability
$P_0$, i.e., the disorder-averaged probability of finding the actual ground state
from the GC technique as a function of $q$ (left panel) and $L$ (right panel),
respectively. These probabilities decay strongly with increasing $q$ and $L$, and
both plots are consistent with an exponential behavior that should be expected when
applying a polynomial-time algorithm to an NP hard problem. Note, however, that the
values of $P_0$ are for GC runs with a single initial configuration that take only
fractions of a second (see the discussion of run times below in
Sec.~\ref{sec:timing}). In real applications one would normally perform runs for many
initial conditions and pick the state of lowest energy. This approach is a generic
method of improving global optimization algorithms
\cite{weigel:06b,khoshbakht:17a}. The success probability of a sequence of $m$ runs
with different initial conditions follows an exponential,
\begin{equation}
  P_s(\{h_i^\alpha\})=1-[1-P_{0}(\{h_i^\alpha\})]^m.
  \label{eq:success}
\end{equation}
Hence for a certain target success probability $P_s$, the
required number of runs follows from
\begin{equation}
m(\{h_i^\alpha\}) = \ln[1-P_s]/\ln[1-P_{0}(\{h_i^\alpha\})],
\end{equation}
where we write $m(\{h_i^\alpha\})$ and $P_{n}(\{h_i^\alpha\})$ to indicate that this
is for a single disorder realization.  With $P_0 = 0.00187$ for $q=3$ and $L=40$
shown in the right panel of Fig.~\ref{fig:success}, for example, using $m=2460$ runs
ensures $P_s = 0.99$ \footnote{Note that Eq.~(\ref{eq:success}) is actually valid at
  the level of individual disorder samples, but it also typically provides a good
  approximation if used at the level of disorder averages \cite{weigel:06b}.}.

In order to compare the performance of GC and PT, we tune the latter via the number
of Monte Carlo steps used to yield the same average success probability (on the same
set of samples) as GC. This can be easily achieved without additional calculations
from the onset times determined in Sec.~\ref{sec:tempering}: the number of steps
$t^\ast$ for all runs is chosen such that the fraction $n(t^\ast)/N_s$ of samples
with $t_0 < t^\ast$ exactly equals the success probability $P_0$ observed for GC,
where $N_s=1536$ is the total number of samples studied. This is illustrated by the data
points for PT also shown in Fig.~\ref{fig:success} that fall on top of the results
for GC.

\begin{figure}[bt!]
  \centering
  \includegraphics[width=.475\columnwidth]{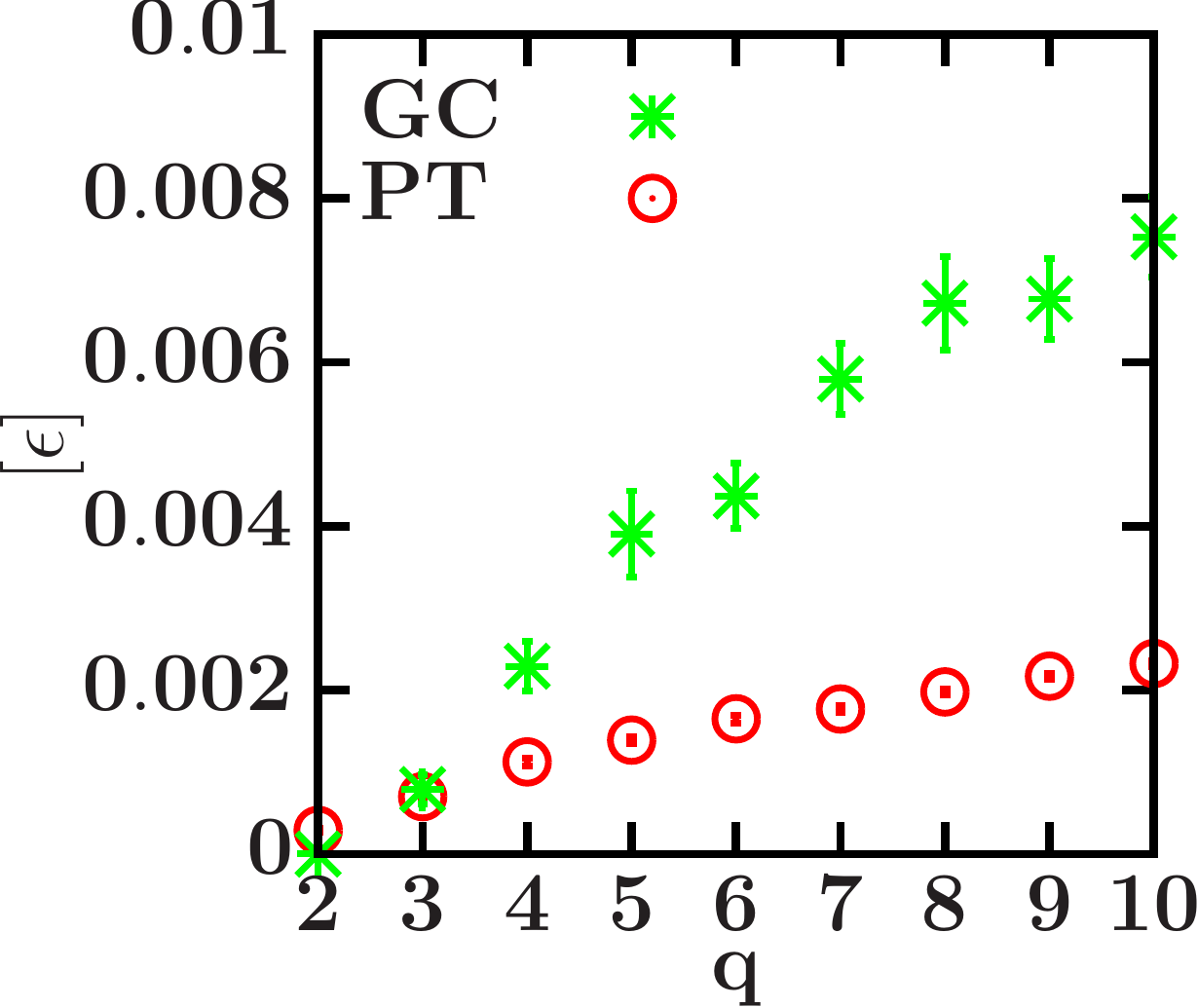}
  \includegraphics[width=.475\columnwidth]{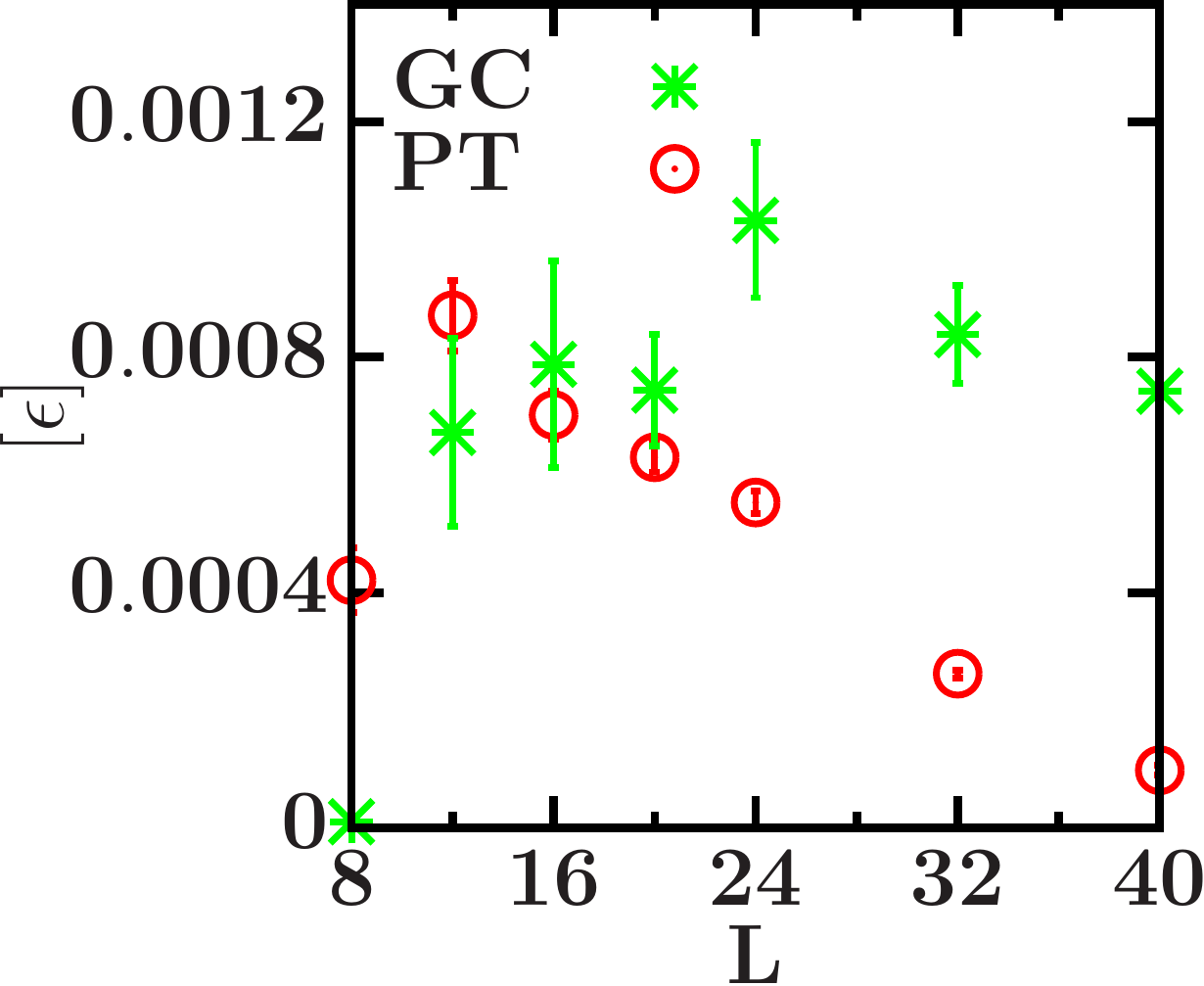}
  \caption{
    Accuracy $[\varepsilon]$ defined in Eq.~(\ref{accu}) of GC and PT
    runs for the $d=2$ RFPM as a function of $q$ (left panel, $L=16$) and $L$
    (right panel, $q=3$), respectively. Both methods are tuned to have the same success
    probabilities, as shown in Fig.~\ref{fig:success}. The data are averaged over
    1536  disorder realizations with $\Delta=1.0$.
  }
    \label{accu_gc}
\end{figure}

\begin{figure}[tb!]
  \centering
  \includegraphics[width=0.475\columnwidth]{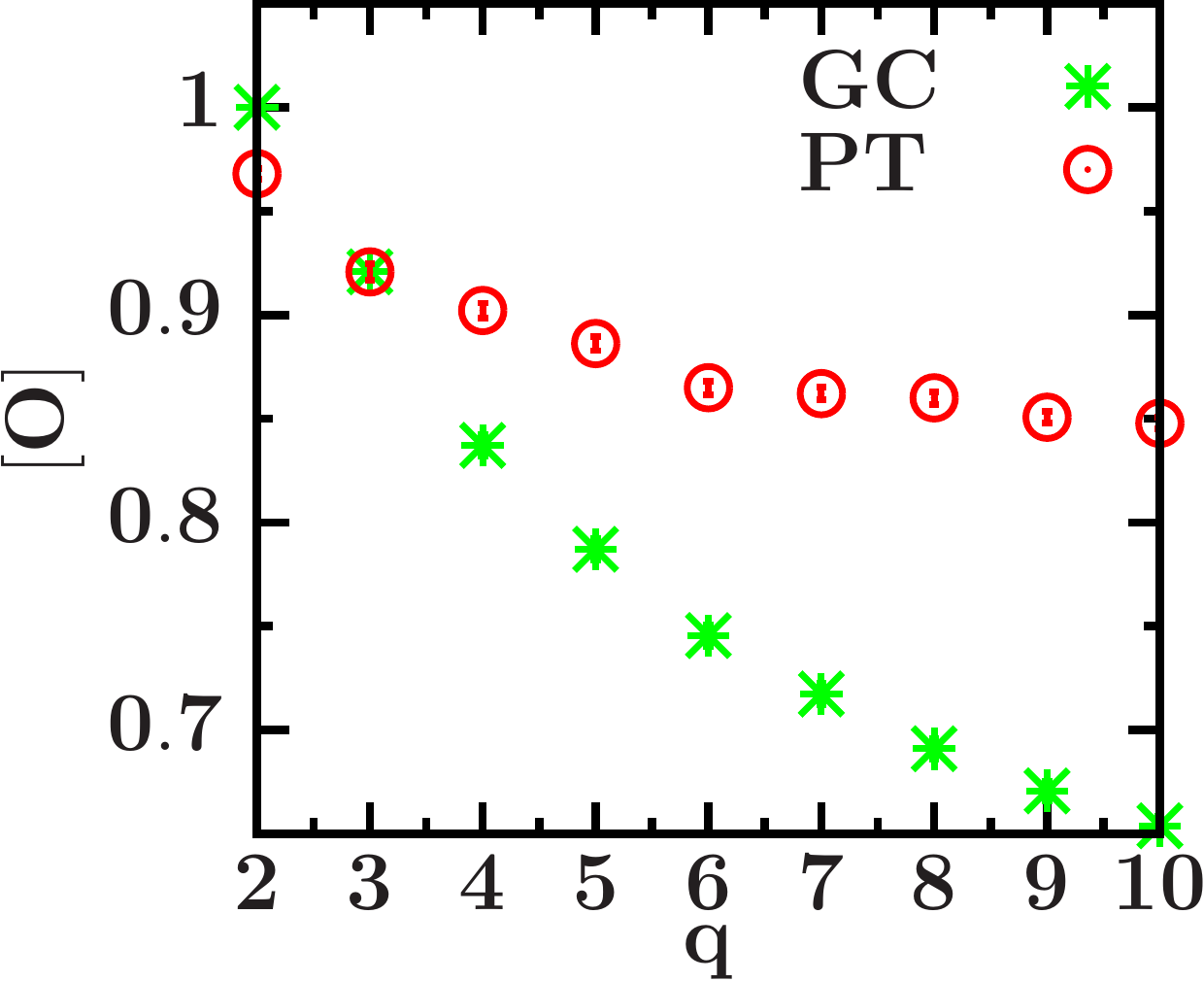}
  \includegraphics[width=0.475\columnwidth]{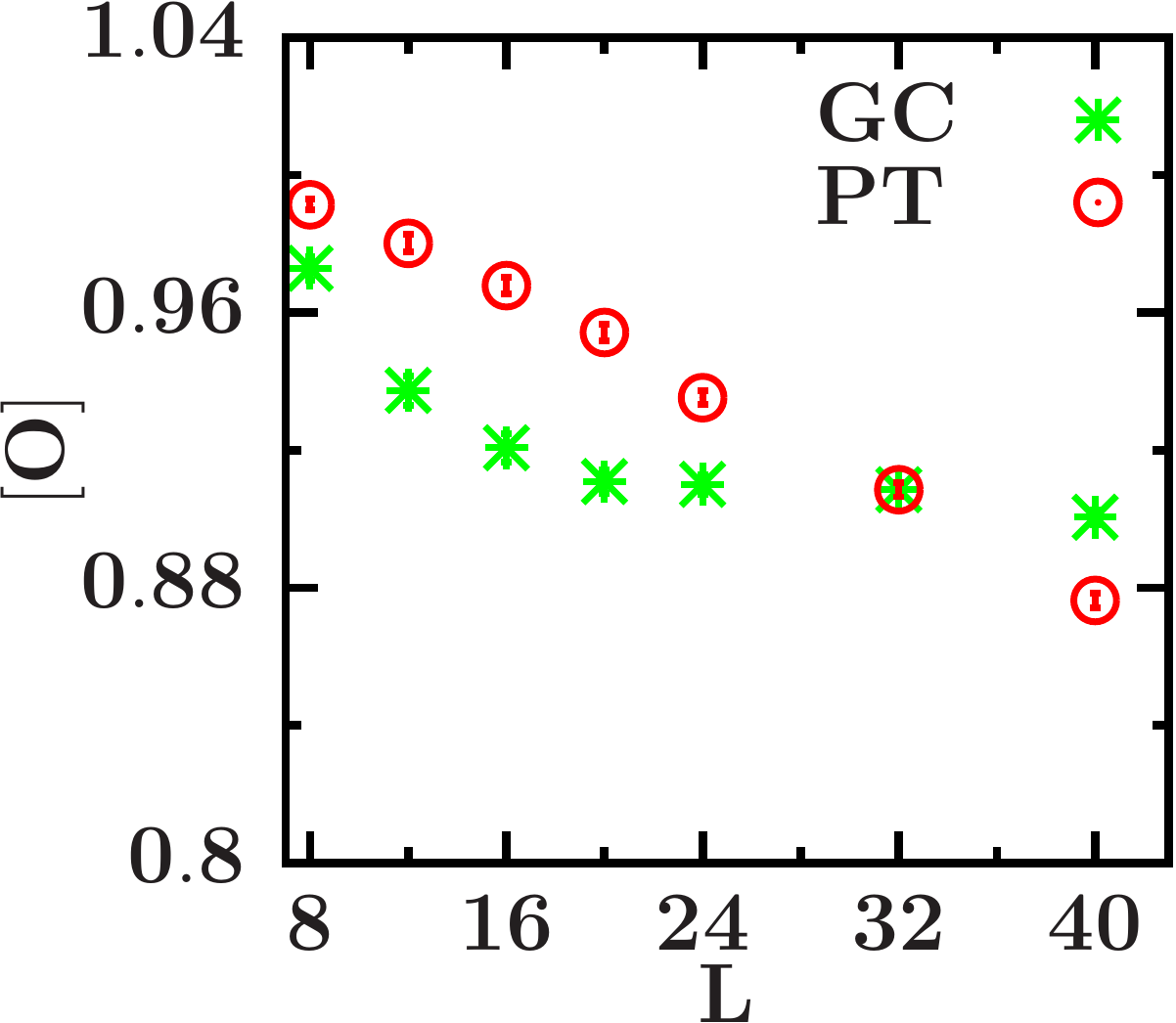}
  \caption{ Average overlap $[O]$ [see Eq.~\eqref{overlap}] of the states returned by
    GC and PT, respectively, with the true ground states for the $d=2$ RFPM as a
    function of $q$ for $L=16$ (left) and $L$ for $q=3$ (right), respectively. The
    averaging is done over 1536 disorder realizations with $\Delta=1.0$.}
  \label{fig:over}
\end{figure}

While the success probabilities of one GC run and the PT simulation with $t^\ast$
steps are identical, this does not imply that both methods find the same states in
case they do {\em not\/} arrive at ground states. To quantify the quality of
approximation in these cases, we consider the relative excess energy of the minimum
energies returned by both algorithms above the ground state,
\begin{equation}
  \label{accu}
  \varepsilon=\frac{E_\mathrm{min}-E_0}{E_0}.
\end{equation}
This quantity, which we call {\em accuracy\/}, is shown in Fig.~\ref{accu_gc} which
reveals that the accuracy at the same success probability is approximately comparable
as a function of $L$ and for $q=3$ in the regime considered, but the approximation
provided by the GC approach appears to more rapidly deteriorate as $q$ is increased
than that of PT. Note that $[\epsilon] = 0$ for GC at $q=2$ as this method finds
exact ground states for the RFIM. We also considered the {\it overlap\/},
\begin{equation}
\label{overlap}
O=\frac{1}{N}\sum_{i=1}^N \delta_{s_i, s_i^{0}}
\end{equation}
of the minimum-energy configurations $\{s_i\}$ found with the true ground states
$\{s_i^{0}\}$. This is shown in Fig.~\ref{fig:over}. As for the accuracy, the overlap
decreases quickly with increasing $q$. For $q=3$, on the other hand, overlaps are
generally high and decrease only moderately with $L$. For the GC approach, there is a
tendency of the decay to settle for $L\gtrsim 32$, promising to provide states of
high similarity to the true ground states even for larger system sizes.

\subsection{Run times and computational complexity}
\label{sec:timing}

\begin{figure}[tb!]
  \centering
  \includegraphics[width=0.95\columnwidth]{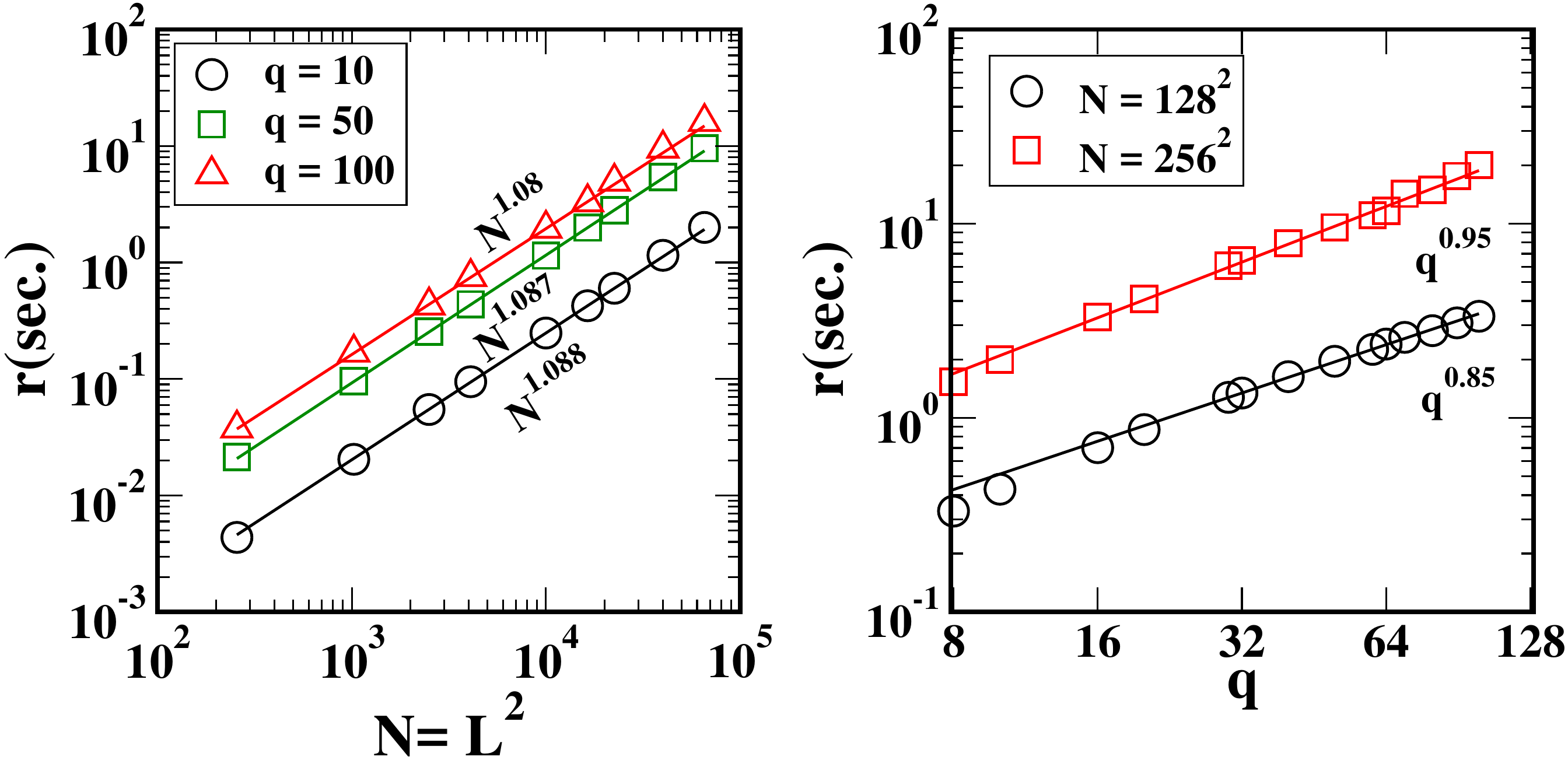}
  \includegraphics[width=0.95\columnwidth]{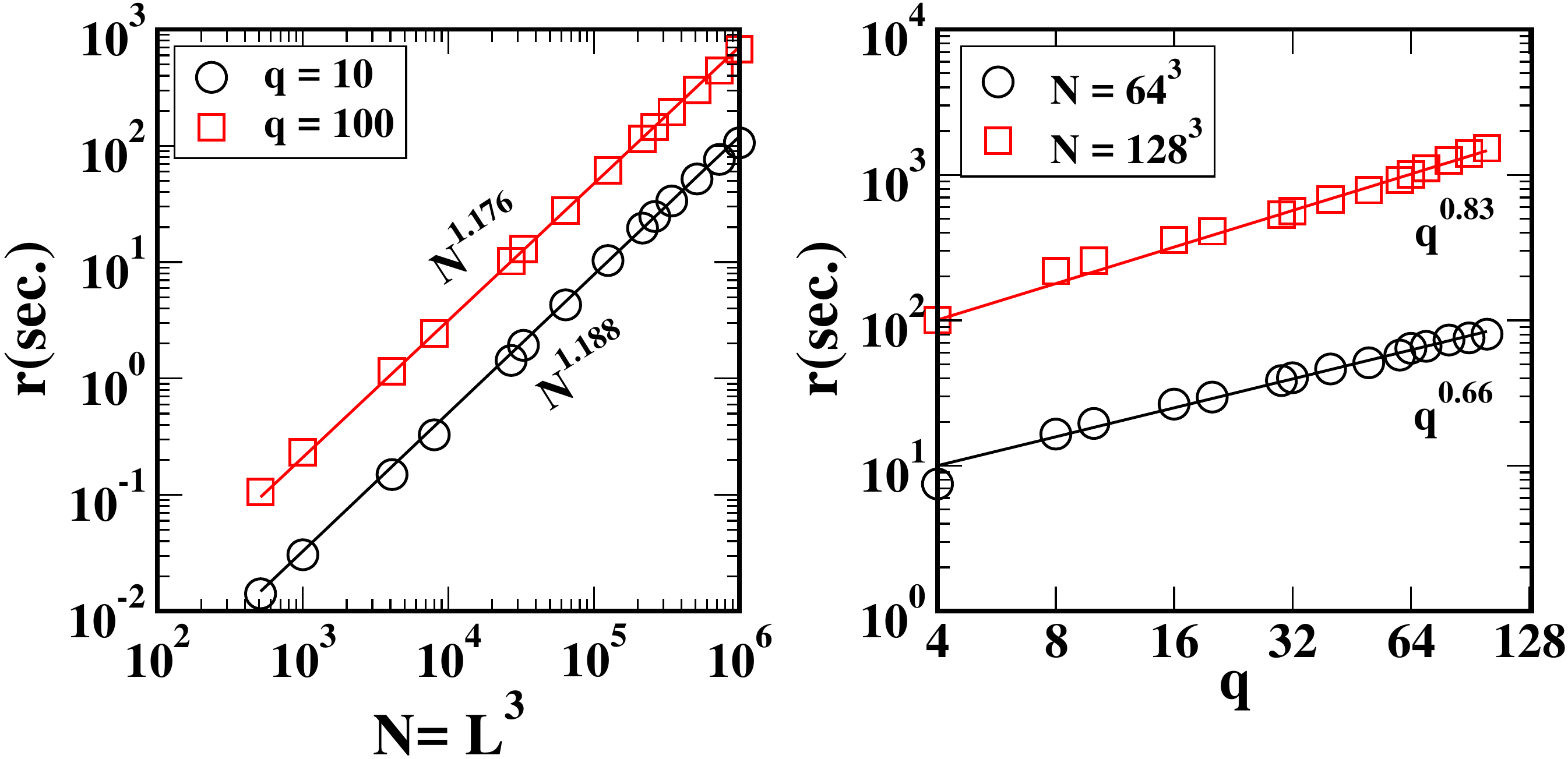}
  \caption{
    Disorder-averaged run time $[r]$ (in CPU sec.) for determining the final state by the
    application of the $\alpha$-expansion GC method to the RFPM in two dimensions
    (top row) and three dimensions (bottom row) as a function of the number of spins
    $N=L^d$ and the number of states $q$, respectively.
    The data are averaged over 1000 disorder realizations with $\Delta=1.0$.
    The solid lines are power-law fits with the specified exponents, and demonstrate
    that the run time is linear in $N$ and approximately linear in $q$.
  }
  \label{run}
\end{figure}

Let us now discuss the time taken by the GC method to find an approximate ground
state of the RFPM. We measure the CPU time $r$ (in seconds) that the
$\alpha$-expansion variant of GC used here takes to reach its final state. We ran our
codes on an IBM cluster with 2.67 GHz Intel Xeon processors. The simulations are
performed for $\Delta=1.0$, and $r$ is averaged over 1000 disorder
samples. Fig.~\ref{run} (top row) shows the run time $[r]$ for the $q$-state RFPM
in $d=2$. We plot $[r]$ as a function of (a) the total number of spins $N=L^2$ for
$q=10$, $50$, $100$; and (b) $q$ for $L=128$, $256$. The solid lines are power-law
fits with the specified exponent. Clearly, $[r]$ is linear in $N$ and $q$ for the
$q$-state RFPM. This is in line with the general discussion of the time complexity of
the method given in Sec.~\ref{sec:gc}. A similar analysis for the RFPM in three
dimensions is summarized in the bottom row of Fig.~\ref{run} which shows that also
in this case the run time is approximately linear with respect to $N$ and $q$.

\begin{figure}[tb!]
  \centering
  \includegraphics[width=0.475\columnwidth]{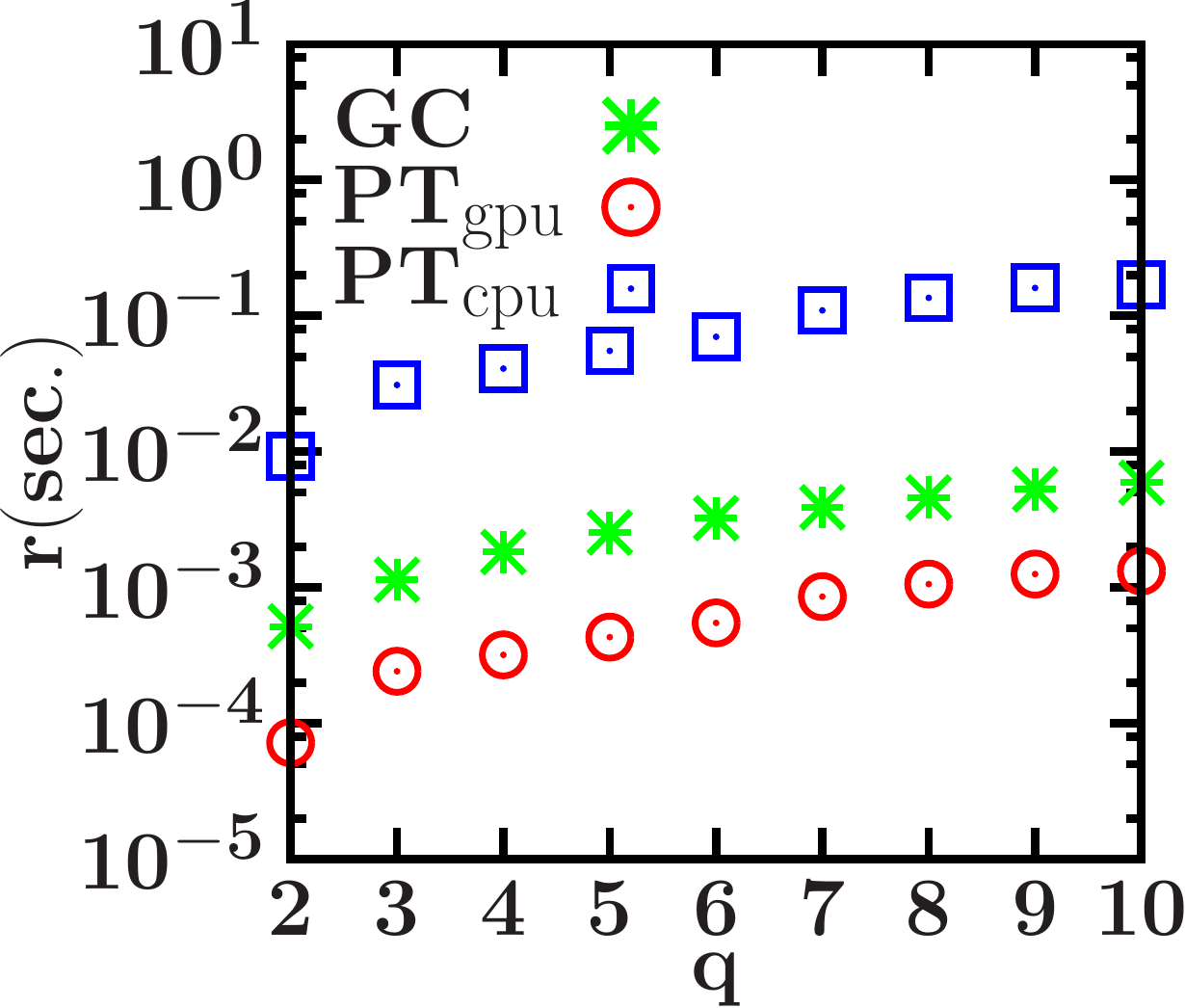}
  \includegraphics[width=0.475\columnwidth]{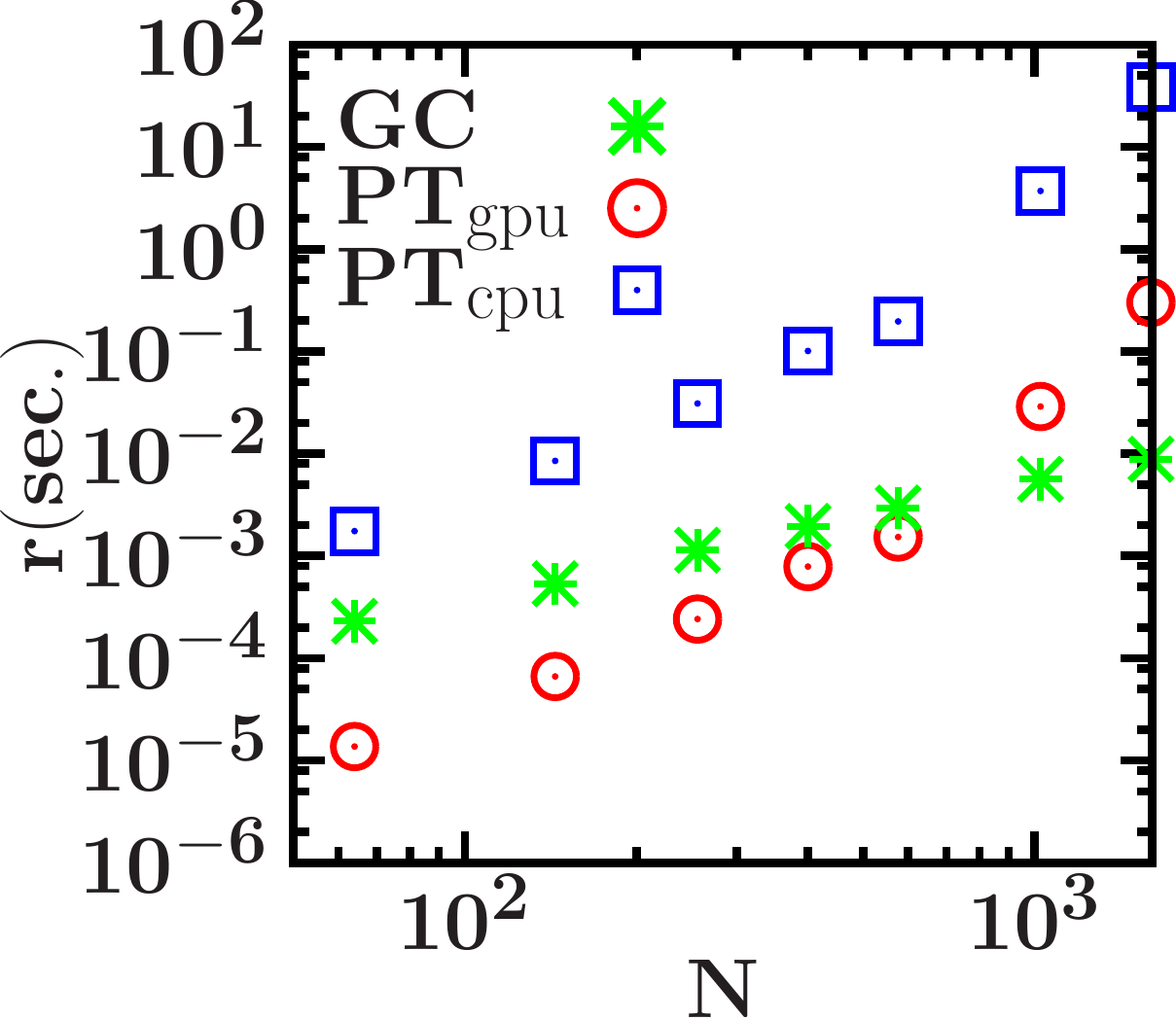}
  \caption{
    Disorder-averaged run time $[r]$ of the GC method for the 2d  RFPM
    as compared to CPU and GPU implementations of the PT method tuned to achieve the
    same success probability as a function of $q$ (left panel, $N=16^2$) and as a
    function of $N=L^2$ (right panel, $q=3$).
  }
  \label{fig:times}
\end{figure}

We finally consider the scaling of run times of the GC and PT techniques with the
latter scaled to achieve the same success probability in finding ground states as the
former. We compare the timings of the GC method to two different implementations of
PT, one regular CPU code and a highly optimized implementation on graphics processing
units (GPUs) \cite{weigel:18,gross:18}. The GPU code is about 128 times faster than
the CPU implementation. The corresponding run times for the two-dimensional RFPM are
shown in Fig.~\ref{fig:times}, using an Nvidia GTX1060 GPU. The times for the GC
approach depend linearly on $q$ and $N=L^2$ to a very good approximation as already
seen above. The CPU variant of PT is always significantly slower that GC at the same
success probability. The GPU code is slightly faster than GC for small systems, but
for larger system sizes the GC approach becomes more favorable as PT shows a clearly
superlinear increase of run times there. For the system sizes probably used in
practical studies that are significantly larger than the sizes $L\le 40$ considered
with quasi-exact ground states here, we expect a substantial advantage for GC over
PT.

\section{Scaling of the breakup length}
\label{sec:breakup}

\begin{figure}[t!]
  \centering
  \includegraphics[width=.95\columnwidth]{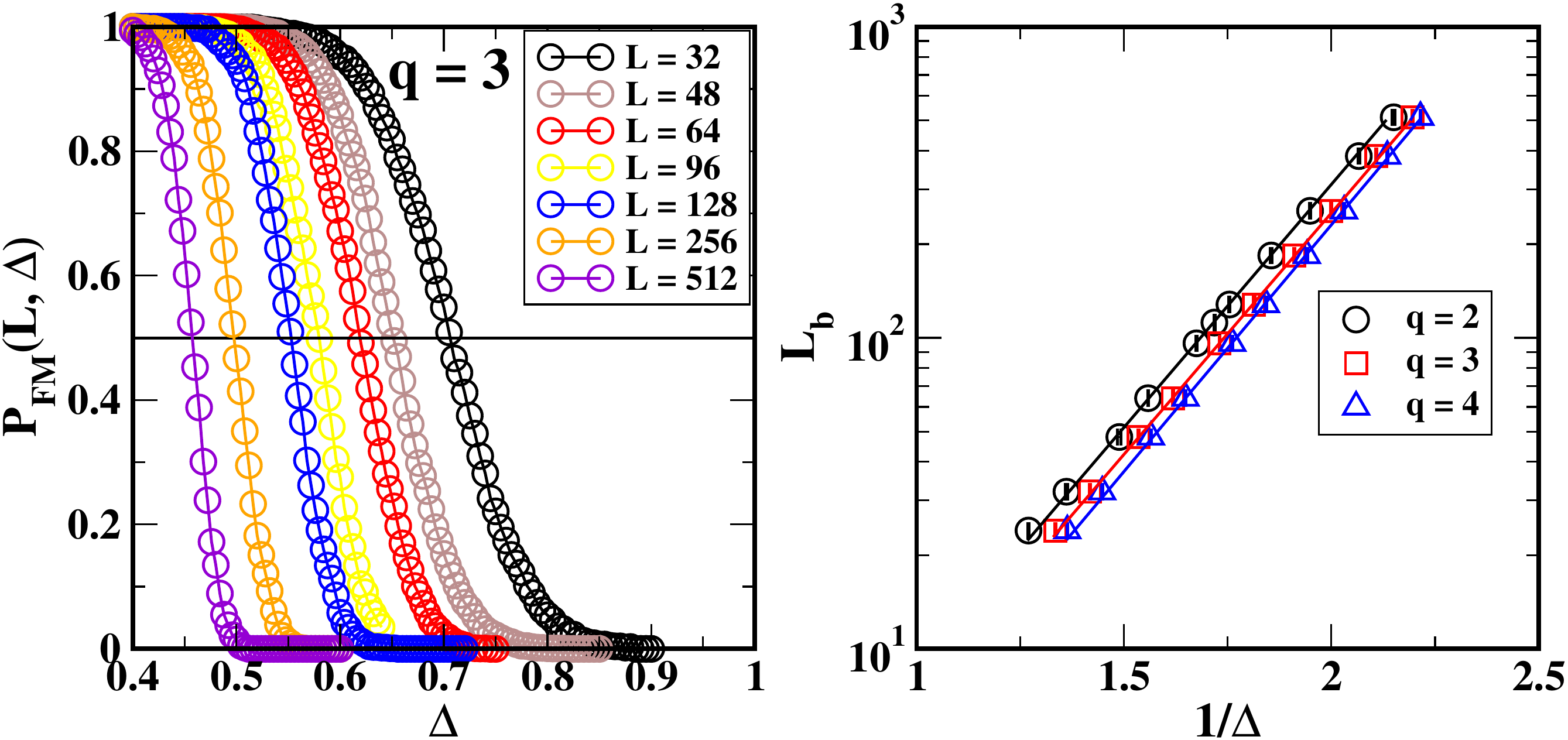}
  \caption {Left panel: Disorder-averaged probability $P_\mathrm{FM}(L,\Delta)$ of
    samples of the 2d $q=3$ RFPM to have purely ferromagnetic ground states.  The
    data are averaged over 10\,000 disorder realizations for $L\le 128$ and 5\,000
    realizations for $L=256$ and $512$. Right panel: The breakup length scale $L_b$,
    defined as the system size $L$ where $P_\mathrm{FM}(L,\Delta) = 0.5$, versus the
    inverse random-field strength $1/\Delta$ for $q=2$, $3$, and $4$. The solid lines
    show fits of the functional form $L_b \sim e^{A/\Delta}$ to the data, where
    $A=3.6 \pm 0.03$.  }
  \label{fig:breakup}
\end{figure}

We finally consider an application of the methods outlined above to exploring the
physical properties of the RFPM in two dimensions. Given the absence of
finite-temperature ordering in the 2d RFIM \cite{binder:83a}, it seems fairly clear
that the RFPM also does not admit order at $T> 0$
\cite{aizenmann:89a,blankschtein:84}. Instead, one expects the presence of ferromagnetic
domains that break up at a length scale $L_b(\Delta)$ similar to what is observed for
the RFIM \cite{binder:83a,seppala:98}.  At very small disorder, the ground state
approaches a purely ferromagnetic state for all but the largest system sizes, while
at large disorder the ground state breaks into domains of $q$ labels. To determine
$L_b$, we follow Ref.~\cite{seppala:98} and count the fraction of samples with a
purely ferromagnetic ground state, defining the probability
$P_\mathrm{FM}(L,\Delta)$. This quantity is shown in Fig.~\ref{fig:breakup}(a) as
determined from GC for $q=3$ and a number of different lattice sizes $L$. The breakup
length $L_b$ can then be defined from the condition $P_\mathrm{FM}(L,\Delta) = 0.5$
\cite{seppala:98}. A plot of $L_b$ vs.\ $1/\Delta$ is shown for the cases $q=2$, $3$,
and $4$ in Fig.~\ref{fig:breakup}(b) using a semi-logarithmic scale. We find that
fits of the simple exponential form
\begin{equation}
  \label{lb}
  L_b \sim \exp(A/\Delta)
\end{equation}
to the data work well, and we arrive at $A=3.6 \pm 0.03$ as a $q$-independent
constant that depends only on the disorder distribution. We note that this scaling is
not consistent with that proposed in Refs.~\cite{binder:83a,seppala:98} for the RFIM,
but it is in line with what was found in numerical simulations of the RFIM in
Ref.~\cite{pre90}. The reason for this discrepancy might be the presence of only a
rather weak curvature in a plot of the type of Fig.~\ref{fig:breakup}(b), and one
might need to go to rather small $\Delta$ to see the asymptotic behavior.

\section{Summary and Discussion}
\label{s4}

The problem of finding the ground state of the $q$-state random field Potts model
(RFPM) corresponds to a multi-terminal flow problem that is known to be NP
hard. Although this model has many physical realizations, the unavailability of
suitable methods has been an impediment in the study of the RFPM. The energy
functions of such complex spin systems have several deep minima separated by
high-energy barriers which grow exponentially with the system size $N$. In this
paper, we have explored the utility of a graph-cut method proposed by Boykov {\em et
  al.} \cite{bvz} for finding approximate ground states of the RFPM. The approach has
the advantage of converging to the final state in polynomial time. However, there is
no guarantee that the states found are ground states for the $q$-state RFPM when
$q \geq 3$. Therefore, it is crucial to benchmark the quality of this approximation.

We have used a carefully tuned set of parallel tempering simulations for creating a
benchmark set of instances for which the ground states are known with an exceedingly
high probability. These allowed to gauge the success probabilities of finding ground
states for the graph-cut method and for short parallel tempering simulations. It is
found that as a function of system size $L$ the quality of the states returned by
graph-cut and parallel tempering techniques is quite similar for small $q$ and system
sizes up to $40\times 40$ spins. For larger systems and $q=3$, there is a tendency of
the graph-cut approach to yield a better approximation. For increasing values of $q$,
on the other hand, the quality of graph-cut results deteriorates rather quickly. The
actual time required for a run of the graph-cut method for small $L$ and different
values of $q$ is much smaller than that of a corresponding parallel tempering run
performed on CPU and comparable to that of a highly efficient GPU implementation of
parallel tempering. For larger system sizes there is a crossover and the graph-cut
approach starts to outperform even the GPU implementation of PT and is likely
asymptotically the most efficient approach. The success probability for the very fast
graph-cut method can be additionally increased by using repeated runs and selecting
the minimum-energy state found among them.
Concerning the comparison of algorithms for the 2d RFPM, we can summarize our
observations as follows:
\begin{enumerate*}[label=(\arabic*)]
\item The PT method guarantees GS in the infinite run-time limit, but the GC method
  gives approximate GS in a very short time $\sim \mathcal{O}(N)$, irrespective of
  the number of states $q$.
\item We find that graph cuts provide an excellent approximation to the ground states
for $q=3$, $4$. The overlap between the ground state and the final states obtained
from the graph cut is very high for smaller $q$ (e.g., $\gtrsim 96\%$ for $q=3$)
and decreases as $q$ is increased.
\item For a fixed value of $q =3$, the overlap between ground states and graph-cut
configuration saturates to a very high value of about $91\%$ for $L\gtrsim 40$.
\end{enumerate*}

The above observations clearly demonstrate that the GC technique is suitable for the
study of the $d=2$ RFPM for lower $q$-values with large system sizes. In particular,
for $q=3$ and $4$, the returned configurations are very close to the exact ground
states. The $q$-state RFPM, though of great physical significance, has received very
little attention due to the unavailability of efficient computational techniques. Our
study sets the stage for investigating this model in particular, and disordered spin
models in general, using methods based on graph cuts. It will be intriguing to make
advances regarding our understanding of the general phase diagram of the RFPM as a
function of $q$ and field strength $\Delta$, in particular for the physically most
relevant three-dimensional case.

\acknowledgments

The authors acknowledge support from the European Commission through the IRSES
network DIONICOS under Contract No.\ PIRSES-GA-2013-612707. MK and VB would like to
acknowledge the support of the Department of Science and Technology, India, through
Grant No. SB/S2/CMP-086/2013.


\begin{thebibliography}{72}%
\makeatletter
\providecommand \@ifxundefined [1]{%
 \@ifx{#1\undefined}
}%
\providecommand \@ifnum [1]{%
 \ifnum #1\expandafter \@firstoftwo
 \else \expandafter \@secondoftwo
 \fi
}%
\providecommand \@ifx [1]{%
 \ifx #1\expandafter \@firstoftwo
 \else \expandafter \@secondoftwo
 \fi
}%
\providecommand \natexlab [1]{#1}%
\providecommand \enquote  [1]{``#1''}%
\providecommand \bibnamefont  [1]{#1}%
\providecommand \bibfnamefont [1]{#1}%
\providecommand \citenamefont [1]{#1}%
\providecommand \href@noop [0]{\@secondoftwo}%
\providecommand \href [0]{\begingroup \@sanitize@url \@href}%
\providecommand \@href[1]{\@@startlink{#1}\@@href}%
\providecommand \@@href[1]{\endgroup#1\@@endlink}%
\providecommand \@sanitize@url [0]{\catcode `\\12\catcode `\$12\catcode
  `\&12\catcode `\#12\catcode `\^12\catcode `\_12\catcode `\%12\relax}%
\providecommand \@@startlink[1]{}%
\providecommand \@@endlink[0]{}%
\providecommand \url  [0]{\begingroup\@sanitize@url \@url }%
\providecommand \@url [1]{\endgroup\@href {#1}{\urlprefix }}%
\providecommand \urlprefix  [0]{URL }%
\providecommand \Eprint [0]{\href }%
\providecommand \doibase [0]{http://dx.doi.org/}%
\providecommand \selectlanguage [0]{\@gobble}%
\providecommand \bibinfo  [0]{\@secondoftwo}%
\providecommand \bibfield  [0]{\@secondoftwo}%
\providecommand \translation [1]{[#1]}%
\providecommand \BibitemOpen [0]{}%
\providecommand \bibitemStop [0]{}%
\providecommand \bibitemNoStop [0]{.\EOS\space}%
\providecommand \EOS [0]{\spacefactor3000\relax}%
\providecommand \BibitemShut  [1]{\csname bibitem#1\endcsname}%
\let\auto@bib@innerbib\@empty
\bibitem [{\citenamefont {Wu}(1982)}]{wu:82a}%
  \BibitemOpen
  \bibfield  {author} {\bibinfo {author} {\bibfnamefont {F.~Y.}\ \bibnamefont
  {Wu}},\ }\href {http://dx.doi.org/10.1103/RevModPhys.54.235} {\bibfield
  {journal} {\bibinfo  {journal} {Rev. Mod. Phys.}\ }\textbf {\bibinfo {volume}
  {54}},\ \bibinfo {pages} {235} (\bibinfo {year} {1982})}\BibitemShut
  {NoStop}%
\bibitem [{\citenamefont {Potts}(1952)}]{potts}%
  \BibitemOpen
  \bibfield  {author} {\bibinfo {author} {\bibfnamefont {R.~B.}\ \bibnamefont
  {Potts}},\ }in\ \href@noop {} {\emph {\bibinfo {booktitle} {Mathematical
  proceedings of the Cambridge philosophical society}}},\ Vol.~\bibinfo
  {volume} {48}\ (\bibinfo {organization} {Cambridge University Press},\
  \bibinfo {year} {1952})\ pp.\ \bibinfo {pages} {106--109}\BibitemShut
  {NoStop}%
\bibitem [{\citenamefont {Binder}\ and\ \citenamefont {Reger}(1992)}]{br92}%
  \BibitemOpen
  \bibfield  {author} {\bibinfo {author} {\bibfnamefont {K.}~\bibnamefont
  {Binder}}\ and\ \bibinfo {author} {\bibfnamefont {J.}~\bibnamefont {Reger}},\
  }\href@noop {} {\bibfield  {journal} {\bibinfo  {journal} {Adv. Phys.}\
  }\textbf {\bibinfo {volume} {41}},\ \bibinfo {pages} {547} (\bibinfo {year}
  {1992})}\BibitemShut {NoStop}%
\bibitem [{\citenamefont {Michel}(1987)}]{michel}%
  \BibitemOpen
  \bibfield  {author} {\bibinfo {author} {\bibfnamefont {K.}~\bibnamefont
  {Michel}},\ }\href@noop {} {\bibfield  {journal} {\bibinfo  {journal} {Phys.
  Rev. B}\ }\textbf {\bibinfo {volume} {35}},\ \bibinfo {pages} {1414}
  (\bibinfo {year} {1987})}\BibitemShut {NoStop}%
\bibitem [{\citenamefont {Aharony}\ \emph {et~al.}(1977)\citenamefont
  {Aharony}, \citenamefont {M{\"u}ller},\ and\ \citenamefont
  {Berlinger}}]{aharony}%
  \BibitemOpen
  \bibfield  {author} {\bibinfo {author} {\bibfnamefont {A.}~\bibnamefont
  {Aharony}}, \bibinfo {author} {\bibfnamefont {K.}~\bibnamefont {M{\"u}ller}},
  \ and\ \bibinfo {author} {\bibfnamefont {W.}~\bibnamefont {Berlinger}},\
  }\href@noop {} {\bibfield  {journal} {\bibinfo  {journal} {Phys. Rev. Lett.}\
  }\textbf {\bibinfo {volume} {38}},\ \bibinfo {pages} {33} (\bibinfo {year}
  {1977})}\BibitemShut {NoStop}%
\bibitem [{\citenamefont {Domany}\ \emph {et~al.}(1982)\citenamefont {Domany},
  \citenamefont {Shnidman},\ and\ \citenamefont {Mukamel}}]{dsm}%
  \BibitemOpen
  \bibfield  {author} {\bibinfo {author} {\bibfnamefont {E.}~\bibnamefont
  {Domany}}, \bibinfo {author} {\bibfnamefont {Y.}~\bibnamefont {Shnidman}}, \
  and\ \bibinfo {author} {\bibfnamefont {D.}~\bibnamefont {Mukamel}},\
  }\href@noop {} {\bibfield  {journal} {\bibinfo  {journal} {J. Phys. C}\
  }\textbf {\bibinfo {volume} {15}},\ \bibinfo {pages} {L495} (\bibinfo {year}
  {1982})}\BibitemShut {NoStop}%
\bibitem [{\citenamefont {Chen}\ \emph {et~al.}(1992)\citenamefont {Chen},
  \citenamefont {Ferrenberg},\ and\ \citenamefont {Landau}}]{chen:92}%
  \BibitemOpen
  \bibfield  {author} {\bibinfo {author} {\bibfnamefont {S.}~\bibnamefont
  {Chen}}, \bibinfo {author} {\bibfnamefont {A.~M.}\ \bibnamefont
  {Ferrenberg}}, \ and\ \bibinfo {author} {\bibfnamefont {D.~P.}\ \bibnamefont
  {Landau}},\ }\href@noop {} {\bibfield  {journal} {\bibinfo  {journal} {Phys.
  Rev. Lett.}\ }\textbf {\bibinfo {volume} {69}},\ \bibinfo {pages} {1213}
  (\bibinfo {year} {1992})}\BibitemShut {NoStop}%
\bibitem [{\citenamefont {Picco}(1997)}]{picco:97}%
  \BibitemOpen
  \bibfield  {author} {\bibinfo {author} {\bibfnamefont {M.}~\bibnamefont
  {Picco}},\ }\href@noop {} {\bibfield  {journal} {\bibinfo  {journal} {Phys.
  Rev. Lett.}\ }\textbf {\bibinfo {volume} {79}},\ \bibinfo {pages} {2998}
  (\bibinfo {year} {1997})}\BibitemShut {NoStop}%
\bibitem [{\citenamefont {Berche}\ and\ \citenamefont
  {Chatelain}(2004)}]{berche:03a}%
  \BibitemOpen
  \bibfield  {author} {\bibinfo {author} {\bibfnamefont {B.}~\bibnamefont
  {Berche}}\ and\ \bibinfo {author} {\bibfnamefont {C.}~\bibnamefont
  {Chatelain}},\ }in\ \href@noop {} {\emph {\bibinfo {booktitle} {Order,
  Disorder And Criticality}}},\ \bibinfo {editor} {edited by\ \bibinfo {editor}
  {\bibfnamefont {Y.}~\bibnamefont {Holovatch}}}\ (\bibinfo  {publisher} {World
  Scientific},\ \bibinfo {address} {Singapore},\ \bibinfo {year} {2004})\ p.\
  \bibinfo {pages} {147}\BibitemShut {NoStop}%
\bibitem [{\citenamefont {Chatelain}\ \emph {et~al.}(2001)\citenamefont
  {Chatelain}, \citenamefont {Berche}, \citenamefont {Janke},\ and\
  \citenamefont {Berche}}]{chatelain:01a}%
  \BibitemOpen
  \bibfield  {author} {\bibinfo {author} {\bibfnamefont {C.}~\bibnamefont
  {Chatelain}}, \bibinfo {author} {\bibfnamefont {B.}~\bibnamefont {Berche}},
  \bibinfo {author} {\bibfnamefont {W.}~\bibnamefont {Janke}}, \ and\ \bibinfo
  {author} {\bibfnamefont {P.~E.}\ \bibnamefont {Berche}},\ }\href@noop {}
  {\bibfield  {journal} {\bibinfo  {journal} {Phys. Rev. E}\ }\textbf {\bibinfo
  {volume} {64}},\ \bibinfo {pages} {036120} (\bibinfo {year}
  {2001})}\BibitemShut {NoStop}%
\bibitem [{\citenamefont {Berche}\ \emph {et~al.}(2002)\citenamefont {Berche},
  \citenamefont {Chatelain}, \citenamefont {Berche},\ and\ \citenamefont
  {Janke}}]{berche:02a}%
  \BibitemOpen
  \bibfield  {author} {\bibinfo {author} {\bibfnamefont {P.~E.}\ \bibnamefont
  {Berche}}, \bibinfo {author} {\bibfnamefont {C.}~\bibnamefont {Chatelain}},
  \bibinfo {author} {\bibfnamefont {B.}~\bibnamefont {Berche}}, \ and\ \bibinfo
  {author} {\bibfnamefont {W.}~\bibnamefont {Janke}},\ }\href@noop {}
  {\bibfield  {journal} {\bibinfo  {journal} {Comput. Phys. Commun.}\ }\textbf
  {\bibinfo {volume} {147}},\ \bibinfo {pages} {427} (\bibinfo {year}
  {2002})}\BibitemShut {NoStop}%
\bibitem [{\citenamefont {Hellmund}\ and\ \citenamefont
  {Janke}(2003)}]{hellmund:03a}%
  \BibitemOpen
  \bibfield  {author} {\bibinfo {author} {\bibfnamefont {M.}~\bibnamefont
  {Hellmund}}\ and\ \bibinfo {author} {\bibfnamefont {W.}~\bibnamefont
  {Janke}},\ }\href@noop {} {\bibfield  {journal} {\bibinfo  {journal} {Phys.
  Rev. E}\ }\textbf {\bibinfo {volume} {67}},\ \bibinfo {pages} {026118}
  (\bibinfo {year} {2003})}\BibitemShut {NoStop}%
\bibitem [{\citenamefont {Chatelain}\ \emph {et~al.}(2005)\citenamefont
  {Chatelain}, \citenamefont {Berche}, \citenamefont {Janke},\ and\
  \citenamefont {Berche}}]{chatelain:05}%
  \BibitemOpen
  \bibfield  {author} {\bibinfo {author} {\bibfnamefont {C.}~\bibnamefont
  {Chatelain}}, \bibinfo {author} {\bibfnamefont {B.}~\bibnamefont {Berche}},
  \bibinfo {author} {\bibfnamefont {W.}~\bibnamefont {Janke}}, \ and\ \bibinfo
  {author} {\bibfnamefont {P.-E.}\ \bibnamefont {Berche}},\ }\href@noop {}
  {\bibfield  {journal} {\bibinfo  {journal} {Nucl. Phys. B}\ }\textbf
  {\bibinfo {volume} {719}},\ \bibinfo {pages} {275} (\bibinfo {year}
  {2005})}\BibitemShut {NoStop}%
\bibitem [{\citenamefont {Blankschtein}\ \emph {et~al.}(1984)\citenamefont
  {Blankschtein}, \citenamefont {Shapir},\ and\ \citenamefont
  {Aharony}}]{blankschtein:84}%
  \BibitemOpen
  \bibfield  {author} {\bibinfo {author} {\bibfnamefont {D.}~\bibnamefont
  {Blankschtein}}, \bibinfo {author} {\bibfnamefont {Y.}~\bibnamefont
  {Shapir}}, \ and\ \bibinfo {author} {\bibfnamefont {A.}~\bibnamefont
  {Aharony}},\ }\href {http://dx.doi.org/10.1103/PhysRevB.29.1263} {\bibfield
  {journal} {\bibinfo  {journal} {Phys. Rev. B}\ }\textbf {\bibinfo {volume}
  {29}},\ \bibinfo {pages} {1263} (\bibinfo {year} {1984})}\BibitemShut
  {NoStop}%
\bibitem [{\citenamefont {Nishimori}(1983)}]{nishimori}%
  \BibitemOpen
  \bibfield  {author} {\bibinfo {author} {\bibfnamefont {H.}~\bibnamefont
  {Nishimori}},\ }\href@noop {} {\bibfield  {journal} {\bibinfo  {journal}
  {Phys. Rev. B}\ }\textbf {\bibinfo {volume} {28}},\ \bibinfo {pages} {4011}
  (\bibinfo {year} {1983})}\BibitemShut {NoStop}%
\bibitem [{\citenamefont {Eichhorn}\ and\ \citenamefont
  {Binder}(1995{\natexlab{a}})}]{eichhorn:95}%
  \BibitemOpen
  \bibfield  {author} {\bibinfo {author} {\bibfnamefont {K.}~\bibnamefont
  {Eichhorn}}\ and\ \bibinfo {author} {\bibfnamefont {K.}~\bibnamefont
  {Binder}},\ }\href {http://dx.doi.org/10.1209/0295-5075/30/6/003} {\bibfield
  {journal} {\bibinfo  {journal} {Europhys. Lett.}\ }\textbf {\bibinfo {volume}
  {30}},\ \bibinfo {pages} {331} (\bibinfo {year}
  {1995}{\natexlab{a}})}\BibitemShut {NoStop}%
\bibitem [{\citenamefont {Eichhorn}\ and\ \citenamefont
  {Binder}(1995{\natexlab{b}})}]{eichhorn:95a}%
  \BibitemOpen
  \bibfield  {author} {\bibinfo {author} {\bibfnamefont {K.}~\bibnamefont
  {Eichhorn}}\ and\ \bibinfo {author} {\bibfnamefont {K.}~\bibnamefont
  {Binder}},\ }\href@noop {} {\bibfield  {journal} {\bibinfo  {journal} {Z.
  Phys. B}\ }\textbf {\bibinfo {volume} {99}},\ \bibinfo {pages} {413}
  (\bibinfo {year} {1995}{\natexlab{b}})}\BibitemShut {NoStop}%
\bibitem [{\citenamefont {Eichhorn}\ and\ \citenamefont
  {Binder}(1996)}]{eichhorn:96}%
  \BibitemOpen
  \bibfield  {author} {\bibinfo {author} {\bibfnamefont {K.}~\bibnamefont
  {Eichhorn}}\ and\ \bibinfo {author} {\bibfnamefont {K.}~\bibnamefont
  {Binder}},\ }\href {http://dx.doi.org/10.1088/0953-8984/8/28/005} {\bibfield
  {journal} {\bibinfo  {journal} {J. Phys.: Condens. Mat.}\ }\textbf {\bibinfo
  {volume} {8}},\ \bibinfo {pages} {5209} (\bibinfo {year} {1996})}\BibitemShut
  {NoStop}%
\bibitem [{\citenamefont {Reed}(1985)}]{reed}%
  \BibitemOpen
  \bibfield  {author} {\bibinfo {author} {\bibfnamefont {P.}~\bibnamefont
  {Reed}},\ }\href@noop {} {\bibfield  {journal} {\bibinfo  {journal} {J. Phys.
  C}\ }\textbf {\bibinfo {volume} {18}},\ \bibinfo {pages} {L615} (\bibinfo
  {year} {1985})}\BibitemShut {NoStop}%
\bibitem [{\citenamefont {Duminil-Copin}\ \emph {et~al.}(2017)\citenamefont
  {Duminil-Copin}, \citenamefont {Sidoravicius},\ and\ \citenamefont
  {Tassion}}]{duminil:15b}%
  \BibitemOpen
  \bibfield  {author} {\bibinfo {author} {\bibfnamefont {H.}~\bibnamefont
  {Duminil-Copin}}, \bibinfo {author} {\bibfnamefont {V.}~\bibnamefont
  {Sidoravicius}}, \ and\ \bibinfo {author} {\bibfnamefont {V.}~\bibnamefont
  {Tassion}},\ }\href@noop {} {\bibfield  {journal} {\bibinfo  {journal}
  {Commun. Math. Phys.}\ }\textbf {\bibinfo {volume} {349}},\ \bibinfo {pages}
  {47} (\bibinfo {year} {2017})}\BibitemShut {NoStop}%
\bibitem [{\citenamefont {Duminil-Copin}\ \emph {et~al.}(2016)\citenamefont
  {Duminil-Copin}, \citenamefont {Gagnebin}, \citenamefont {Harel},
  \citenamefont {Manolescu},\ and\ \citenamefont {Tassion}}]{duminil:16}%
  \BibitemOpen
  \bibfield  {author} {\bibinfo {author} {\bibfnamefont {H.}~\bibnamefont
  {Duminil-Copin}}, \bibinfo {author} {\bibfnamefont {M.}~\bibnamefont
  {Gagnebin}}, \bibinfo {author} {\bibfnamefont {M.}~\bibnamefont {Harel}},
  \bibinfo {author} {\bibfnamefont {I.}~\bibnamefont {Manolescu}}, \ and\
  \bibinfo {author} {\bibfnamefont {V.}~\bibnamefont {Tassion}},\ }\href@noop
{} {\enquote {\bibinfo {title} {Discontinuity of the phase transition for the
      planar random-cluster and {P}otts models with $q>4$},}\ } \bibinfo {note} {preprint
  arXiv:1611.09877}  (\bibinfo {year}
{2016})\BibitemShut {NoStop}%
\bibitem [{\citenamefont {Hartmann}(2005)}]{hartmann:05}%
  \BibitemOpen
  \bibfield  {author} {\bibinfo {author} {\bibfnamefont {A.~K.}\ \bibnamefont
  {Hartmann}},\ }\href {http://prl.aps.org/abstract/PRL/v94/i5/e050601}
  {\bibfield  {journal} {\bibinfo  {journal} {Phys. Rev. Lett.}\ }\textbf
  {\bibinfo {volume} {94}},\ \bibinfo {pages} {050601} (\bibinfo {year}
  {2005})}\BibitemShut {NoStop}%
\bibitem [{\citenamefont {Cardy}(1999)}]{cardy:99a}%
  \BibitemOpen
  \bibfield  {author} {\bibinfo {author} {\bibfnamefont {J.~L.}\ \bibnamefont
  {Cardy}},\ }\href {http://dx.doi.org/10.1016/S0378-4371(98)00489-0}
  {\bibfield  {journal} {\bibinfo  {journal} {Physica A}\ }\textbf {\bibinfo
  {volume} {263}},\ \bibinfo {pages} {215} (\bibinfo {year}
  {1999})}\BibitemShut {NoStop}%
\bibitem [{\citenamefont {Aizenman}\ and\ \citenamefont
  {Wehr}(1989)}]{aizenmann:89a}%
  \BibitemOpen
  \bibfield  {author} {\bibinfo {author} {\bibfnamefont {M.}~\bibnamefont
  {Aizenman}}\ and\ \bibinfo {author} {\bibfnamefont {J.}~\bibnamefont
  {Wehr}},\ }\href {http://dx.doi.org/10.1103/PhysRevLett.62.2503} {\bibfield
  {journal} {\bibinfo  {journal} {Phys. Rev. Lett.}\ }\textbf {\bibinfo
  {volume} {62}},\ \bibinfo {pages} {2503} (\bibinfo {year}
  {1989})}\BibitemShut {NoStop}%
\bibitem [{\citenamefont {Binder}(1983{\natexlab{a}})}]{binder:83}%
  \BibitemOpen
  \bibfield  {author} {\bibinfo {author} {\bibfnamefont {K.}~\bibnamefont
  {Binder}},\ }\href {http://dx.doi.org/10.1007/BF01470045} {\bibfield
  {journal} {\bibinfo  {journal} {Z. Phys. B}\ }\textbf {\bibinfo {volume}
  {50}},\ \bibinfo {pages} {343} (\bibinfo {year}
  {1983}{\natexlab{a}})}\BibitemShut {NoStop}%
\bibitem [{\citenamefont {Hukushima}\ and\ \citenamefont
  {Nemoto}(1996)}]{hukushima:96a}%
  \BibitemOpen
  \bibfield  {author} {\bibinfo {author} {\bibfnamefont {K.}~\bibnamefont
  {Hukushima}}\ and\ \bibinfo {author} {\bibfnamefont {K.}~\bibnamefont
  {Nemoto}},\ }\href@noop {} {\bibfield  {journal} {\bibinfo  {journal} {J.
  Phys. Soc. Jpn.}\ }\textbf {\bibinfo {volume} {65}},\ \bibinfo {pages} {1604}
  (\bibinfo {year} {1996})}\BibitemShut {NoStop}%
\bibitem [{\citenamefont {Nattermann}(1997)}]{nattermann:97}%
  \BibitemOpen
  \bibfield  {author} {\bibinfo {author} {\bibfnamefont {T.}~\bibnamefont
  {Nattermann}},\ }in\ \href {http://arxiv.org/abs/cond-mat/9705295} {\emph
  {\bibinfo {booktitle} {Spin Glasses and Random Fields}}},\ \bibinfo {editor}
  {edited by\ \bibinfo {editor} {\bibfnamefont {A.~P.}\ \bibnamefont {Young}}}\
  (\bibinfo  {publisher} {World Scientific},\ \bibinfo {year} {1997})\ p.\
  \bibinfo {pages} {277},\ \Eprint {http://arxiv.org/abs/cond-mat/9705295}
  {arXiv:cond-mat/9705295} \BibitemShut {NoStop}%
\bibitem [{\citenamefont {Fytas}\ and\ \citenamefont
  {Mart\'{\i}n-Mayor}(2013)}]{fytas:13}%
  \BibitemOpen
  \bibfield  {author} {\bibinfo {author} {\bibfnamefont {N.~G.}\ \bibnamefont
  {Fytas}}\ and\ \bibinfo {author} {\bibfnamefont {V.}~\bibnamefont
  {Mart\'{\i}n-Mayor}},\ }\href
  {http://dx.doi.org/10.1103/PhysRevLett.110.227201} {\bibfield  {journal}
  {\bibinfo  {journal} {Phys. Rev. Lett.}\ }\textbf {\bibinfo {volume} {110}},\
  \bibinfo {pages} {227201} (\bibinfo {year} {2013})}\BibitemShut {NoStop}%
\bibitem [{\citenamefont {Fytas}\ \emph {et~al.}(2016)\citenamefont {Fytas},
  \citenamefont {Mart\'{\i}n-Mayor}, \citenamefont {Picco},\ and\ \citenamefont
  {Sourlas}}]{fytas:16}%
  \BibitemOpen
  \bibfield  {author} {\bibinfo {author} {\bibfnamefont {N.~G.}\ \bibnamefont
  {Fytas}}, \bibinfo {author} {\bibfnamefont {V.}~\bibnamefont
  {Mart\'{\i}n-Mayor}}, \bibinfo {author} {\bibfnamefont {M.}~\bibnamefont
  {Picco}}, \ and\ \bibinfo {author} {\bibfnamefont {N.}~\bibnamefont
  {Sourlas}},\ }\href@noop {} {\bibfield  {journal} {\bibinfo  {journal} {Phys.
  Rev. Lett.}\ }\textbf {\bibinfo {volume} {116}},\ \bibinfo {pages} {227201}
  (\bibinfo {year} {2016})}\BibitemShut {NoStop}%
\bibitem [{\citenamefont {Parisi}\ and\ \citenamefont
  {Sourlas}(1979)}]{parisi:79a}%
  \BibitemOpen
  \bibfield  {author} {\bibinfo {author} {\bibfnamefont {G.}~\bibnamefont
  {Parisi}}\ and\ \bibinfo {author} {\bibfnamefont {N.}~\bibnamefont
  {Sourlas}},\ }\href@noop {} {\bibfield  {journal} {\bibinfo  {journal} {Phys.
  Rev. Lett.}\ }\textbf {\bibinfo {volume} {43}},\ \bibinfo {pages} {744}
  (\bibinfo {year} {1979})}\BibitemShut {NoStop}%
\bibitem [{\citenamefont {Bray}\ and\ \citenamefont {Moore}(1985)}]{bray:85a}%
  \BibitemOpen
  \bibfield  {author} {\bibinfo {author} {\bibfnamefont {A.~J.}\ \bibnamefont
  {Bray}}\ and\ \bibinfo {author} {\bibfnamefont {M.~A.}\ \bibnamefont
  {Moore}},\ }\href@noop {} {\bibfield  {journal} {\bibinfo  {journal} {J.
  Phys. C}\ }\textbf {\bibinfo {volume} {18}} (\bibinfo {year}
  {1985})}\BibitemShut {NoStop}%
\bibitem [{\citenamefont {Angl\`{e}s~d'Auriac}\ \emph
  {et~al.}(1985)\citenamefont {Angl\`{e}s~d'Auriac}, \citenamefont
  {Preissmann},\ and\ \citenamefont {Rammal}}]{dauriac:85}%
  \BibitemOpen
  \bibfield  {author} {\bibinfo {author} {\bibfnamefont {J.~C.}\ \bibnamefont
  {Angl\`{e}s~d'Auriac}}, \bibinfo {author} {\bibfnamefont {M.}~\bibnamefont
  {Preissmann}}, \ and\ \bibinfo {author} {\bibfnamefont {R.}~\bibnamefont
  {Rammal}},\ }\href@noop {} {\bibfield  {journal} {\bibinfo  {journal} {J.
  Physique Lett.}\ }\textbf {\bibinfo {volume} {46}},\ \bibinfo {pages} {L173}
  (\bibinfo {year} {1985})}\BibitemShut {NoStop}%
\bibitem [{\citenamefont {Ford~Jr}\ and\ \citenamefont
  {Fulkerson}(2015)}]{ff5}%
  \BibitemOpen
  \bibfield  {author} {\bibinfo {author} {\bibfnamefont {L.~R.}\ \bibnamefont
  {Ford~Jr}}\ and\ \bibinfo {author} {\bibfnamefont {D.~R.}\ \bibnamefont
  {Fulkerson}},\ }\href@noop {} {\emph {\bibinfo {title} {Flows in Networks}}}\
  (\bibinfo  {publisher} {Princeton University Press},\ \bibinfo {year}
  {2015})\BibitemShut {NoStop}%
\bibitem [{\citenamefont {Goldberg}\ and\ \citenamefont {Tarjan}(1988)}]{gt5}%
  \BibitemOpen
  \bibfield  {author} {\bibinfo {author} {\bibfnamefont {A.~V.}\ \bibnamefont
  {Goldberg}}\ and\ \bibinfo {author} {\bibfnamefont {R.~E.}\ \bibnamefont
  {Tarjan}},\ }\href@noop {} {\bibfield  {journal} {\bibinfo  {journal} {J.
  ACM}\ }\textbf {\bibinfo {volume} {35}},\ \bibinfo {pages} {921} (\bibinfo
  {year} {1988})}\BibitemShut {NoStop}%
\bibitem [{\citenamefont {Boykov}\ and\ \citenamefont
  {Kolmogorov}(2004)}]{kolmogorov:04}%
  \BibitemOpen
  \bibfield  {author} {\bibinfo {author} {\bibfnamefont {Y.}~\bibnamefont
  {Boykov}}\ and\ \bibinfo {author} {\bibfnamefont {V.}~\bibnamefont
  {Kolmogorov}},\ }\href {http://dx.doi.org/10.1109/tpami.2004.60} {\bibfield
  {journal} {\bibinfo  {journal} {IEEE T. Pattern Anal.}\ }\textbf {\bibinfo
  {volume} {26}},\ \bibinfo {pages} {1124} (\bibinfo {year}
  {2004})}\BibitemShut {NoStop}%
\bibitem [{\citenamefont {Shrivastav}\ \emph {et~al.}(2011)\citenamefont
  {Shrivastav}, \citenamefont {Krishnamoorthy}, \citenamefont {Banerjee},\ and\
  \citenamefont {Puri}}]{epl96}%
  \BibitemOpen
  \bibfield  {author} {\bibinfo {author} {\bibfnamefont {G.~P.}\ \bibnamefont
  {Shrivastav}}, \bibinfo {author} {\bibfnamefont {S.}~\bibnamefont
  {Krishnamoorthy}}, \bibinfo {author} {\bibfnamefont {V.}~\bibnamefont
  {Banerjee}}, \ and\ \bibinfo {author} {\bibfnamefont {S.}~\bibnamefont
  {Puri}},\ }\href@noop {} {\bibfield  {journal} {\bibinfo  {journal}
  {Europhys. Lett.}\ }\textbf {\bibinfo {volume} {96}},\ \bibinfo {pages}
  {36003} (\bibinfo {year} {2011})}\BibitemShut {NoStop}%
\bibitem [{\citenamefont {Shrivastav}\ \emph
  {et~al.}(2014{\natexlab{a}})\citenamefont {Shrivastav}, \citenamefont
  {Kumar}, \citenamefont {Banerjee},\ and\ \citenamefont {Puri}}]{pre90}%
  \BibitemOpen
  \bibfield  {author} {\bibinfo {author} {\bibfnamefont {G.~P.}\ \bibnamefont
  {Shrivastav}}, \bibinfo {author} {\bibfnamefont {M.}~\bibnamefont {Kumar}},
  \bibinfo {author} {\bibfnamefont {V.}~\bibnamefont {Banerjee}}, \ and\
  \bibinfo {author} {\bibfnamefont {S.}~\bibnamefont {Puri}},\ }\href@noop {}
  {\bibfield  {journal} {\bibinfo  {journal} {Phys. Rev. E}\ }\textbf {\bibinfo
  {volume} {90}},\ \bibinfo {pages} {032140} (\bibinfo {year}
  {2014}{\natexlab{a}})}\BibitemShut {NoStop}%
\bibitem [{\citenamefont {Shrivastav}\ \emph
  {et~al.}(2014{\natexlab{b}})\citenamefont {Shrivastav}, \citenamefont
  {Banerjee},\ and\ \citenamefont {Puri}}]{epj37}%
  \BibitemOpen
  \bibfield  {author} {\bibinfo {author} {\bibfnamefont {G.~P.}\ \bibnamefont
  {Shrivastav}}, \bibinfo {author} {\bibfnamefont {V.}~\bibnamefont
  {Banerjee}}, \ and\ \bibinfo {author} {\bibfnamefont {S.}~\bibnamefont
  {Puri}},\ }\href@noop {} {\bibfield  {journal} {\bibinfo  {journal} {Eur.
  Phys. J. E}\ }\textbf {\bibinfo {volume} {37}},\ \bibinfo {pages} {98}
  (\bibinfo {year} {2014}{\natexlab{b}})}\BibitemShut {NoStop}%
\bibitem [{\citenamefont {Banerjee}\ \emph {et~al.}(2014)\citenamefont
  {Banerjee}, \citenamefont {Puri},\ and\ \citenamefont {Shrivastav}}]{ijp88}%
  \BibitemOpen
  \bibfield  {author} {\bibinfo {author} {\bibfnamefont {V.}~\bibnamefont
  {Banerjee}}, \bibinfo {author} {\bibfnamefont {S.}~\bibnamefont {Puri}}, \
  and\ \bibinfo {author} {\bibfnamefont {G.~P.}\ \bibnamefont {Shrivastav}},\
  }\href@noop {} {\bibfield  {journal} {\bibinfo  {journal} {Indian J. Phys.}\
  }\textbf {\bibinfo {volume} {88}},\ \bibinfo {pages} {1005} (\bibinfo {year}
  {2014})}\BibitemShut {NoStop}%
\bibitem [{\citenamefont {Stevenson}\ and\ \citenamefont
  {Weigel}(2011{\natexlab{a}})}]{stevenson:11}%
  \BibitemOpen
  \bibfield  {author} {\bibinfo {author} {\bibfnamefont {J.~D.}\ \bibnamefont
  {Stevenson}}\ and\ \bibinfo {author} {\bibfnamefont {M.}~\bibnamefont
  {Weigel}},\ }\href {http://arxiv.org/abs/1010.5973} {\bibfield  {journal}
  {\bibinfo  {journal} {Europhys. Lett.}\ }\textbf {\bibinfo {volume} {95}},\
  \bibinfo {pages} {40001} (\bibinfo {year} {2011}{\natexlab{a}})}
\BibitemShut {NoStop}%
\bibitem [{\citenamefont {Stevenson}\ and\ \citenamefont
  {Weigel}(2011{\natexlab{b}})}]{stevenson:11a}%
  \BibitemOpen
  \bibfield  {author} {\bibinfo {author} {\bibfnamefont {J.~D.}\ \bibnamefont
  {Stevenson}}\ and\ \bibinfo {author} {\bibfnamefont {M.}~\bibnamefont
  {Weigel}},\ }\href {http://dx.doi.org/10.1016/j.cpc.2010.11.028} {\bibfield
  {journal} {\bibinfo  {journal} {Comput. Phys. Commun.}\ }\textbf {\bibinfo
  {volume} {182}},\ \bibinfo {pages} {1879} (\bibinfo {year}
  {2011}{\natexlab{b}})}\BibitemShut {NoStop}%
\bibitem [{\citenamefont {Boykov}\ \emph {et~al.}(2001)\citenamefont {Boykov},
  \citenamefont {Veksler},\ and\ \citenamefont {Zabih}}]{bvz}%
  \BibitemOpen
  \bibfield  {author} {\bibinfo {author} {\bibfnamefont {Y.}~\bibnamefont
  {Boykov}}, \bibinfo {author} {\bibfnamefont {O.}~\bibnamefont {Veksler}}, \
  and\ \bibinfo {author} {\bibfnamefont {R.}~\bibnamefont {Zabih}},\
  }\href@noop {} {\bibfield  {journal} {\bibinfo  {journal} {IEEE T. Pattern
  Anal.}\ }\textbf {\bibinfo {volume} {23}},\ \bibinfo {pages} {1222} (\bibinfo
  {year} {2001})}\BibitemShut {NoStop}%
\bibitem [{\citenamefont {Goldschmidt}\ and\ \citenamefont {Xu}(1985)}]{gx}%
  \BibitemOpen
  \bibfield  {author} {\bibinfo {author} {\bibfnamefont {Y.~Y.}\ \bibnamefont
  {Goldschmidt}}\ and\ \bibinfo {author} {\bibfnamefont {G.}~\bibnamefont
  {Xu}},\ }\href@noop {} {\bibfield  {journal} {\bibinfo  {journal} {Phys. Rev.
  B}\ }\textbf {\bibinfo {volume} {32}},\ \bibinfo {pages} {1876} (\bibinfo
  {year} {1985})}\BibitemShut {NoStop}%
\bibitem [{\citenamefont {Weigel}\ and\ \citenamefont
  {Gingras}(2006)}]{weigel:05f}%
  \BibitemOpen
  \bibfield  {author} {\bibinfo {author} {\bibfnamefont {M.}~\bibnamefont
  {Weigel}}\ and\ \bibinfo {author} {\bibfnamefont {M.~J.~P.}\ \bibnamefont
  {Gingras}},\ }\href@noop {} {\bibfield  {journal} {\bibinfo  {journal} {Phys.
  Rev. Lett.}\ }\textbf {\bibinfo {volume} {96}},\ \bibinfo {pages} {097206}
  (\bibinfo {year} {2006})}\BibitemShut {NoStop}%
\bibitem [{\citenamefont {Weigel}(2007)}]{weigel:06b}%
  \BibitemOpen
  \bibfield  {author} {\bibinfo {author} {\bibfnamefont {M.}~\bibnamefont
  {Weigel}},\ }\href {http://dx.doi.org/10.1103/PhysRevE.76.066706} {\bibfield
  {journal} {\bibinfo  {journal} {Phys. Rev. E}\ }\textbf {\bibinfo {volume}
  {76}},\ \bibinfo {pages} {066706} (\bibinfo {year} {2007})}\BibitemShut
  {NoStop}%
\bibitem [{\citenamefont {De~Simone}\ \emph {et~al.}(1995)\citenamefont
  {De~Simone}, \citenamefont {Diehl}, \citenamefont {J{\"{u}}nger},
  \citenamefont {Mutzel}, \citenamefont {Reinelt},\ and\ \citenamefont
  {Rinaldi}}]{simone:95}%
  \BibitemOpen
  \bibfield  {author} {\bibinfo {author} {\bibfnamefont {C.}~\bibnamefont
  {De~Simone}}, \bibinfo {author} {\bibfnamefont {M.}~\bibnamefont {Diehl}},
  \bibinfo {author} {\bibfnamefont {M.}~\bibnamefont {J{\"{u}}nger}}, \bibinfo
  {author} {\bibfnamefont {P.}~\bibnamefont {Mutzel}}, \bibinfo {author}
  {\bibfnamefont {G.}~\bibnamefont {Reinelt}}, \ and\ \bibinfo {author}
  {\bibfnamefont {G.}~\bibnamefont {Rinaldi}},\ }\href@noop {} {\bibfield
  {journal} {\bibinfo  {journal} {J. Stat. Phys.}\ }\textbf {\bibinfo {volume}
  {80}},\ \bibinfo {pages} {487} (\bibinfo {year} {1995})}\BibitemShut
  {NoStop}%
\bibitem [{\citenamefont {Kirkpatrick}\ \emph {et~al.}(1983)\citenamefont
  {Kirkpatrick}, \citenamefont {Gelatt},\ and\ \citenamefont
  {Vecchi}}]{kirkpatrick:83}%
  \BibitemOpen
  \bibfield  {author} {\bibinfo {author} {\bibfnamefont {S.}~\bibnamefont
  {Kirkpatrick}}, \bibinfo {author} {\bibfnamefont {C.~D.}\ \bibnamefont
  {Gelatt}}, \ and\ \bibinfo {author} {\bibfnamefont {M.~P.}\ \bibnamefont
  {Vecchi}},\ }\href@noop {} {\bibfield  {journal} {\bibinfo  {journal}
  {Science}\ }\textbf {\bibinfo {volume} {220}},\ \bibinfo {pages} {671}
  (\bibinfo {year} {1983})}\BibitemShut {NoStop}%
\bibitem [{\citenamefont {Geyer}(1991)}]{geyer:91}%
  \BibitemOpen
  \bibfield  {author} {\bibinfo {author} {\bibfnamefont {C.~J.}\ \bibnamefont
  {Geyer}},\ }in\ \href@noop {} {\emph {\bibinfo {booktitle} {Computing Science
  and Statistics: Proceedings of the 23rd Symposium on the Interface}}}\
  (\bibinfo  {publisher} {American Statistical Association},\ \bibinfo
  {address} {New York},\ \bibinfo {year} {1991})\ p.\ \bibinfo {pages}
  {156}\BibitemShut {NoStop}%
\bibitem [{\citenamefont {Wang}\ \emph {et~al.}(2015)\citenamefont {Wang},
  \citenamefont {Machta},\ and\ \citenamefont {Katzgraber}}]{wang:15}%
  \BibitemOpen
  \bibfield  {author} {\bibinfo {author} {\bibfnamefont {W.}~\bibnamefont
  {Wang}}, \bibinfo {author} {\bibfnamefont {J.}~\bibnamefont {Machta}}, \ and\
  \bibinfo {author} {\bibfnamefont {H.~G.}\ \bibnamefont {Katzgraber}},\
  }\href@noop {} {\bibfield  {journal} {\bibinfo  {journal} {Phys. Rev. E}\
  }\textbf {\bibinfo {volume} {92}},\ \bibinfo {pages} {013303} (\bibinfo
  {year} {2015})}\BibitemShut {NoStop}%
\bibitem [{Note1()}]{Note1}%
  \BibitemOpen
  \bibinfo {note} {But note that population annealing \cite
  {hukushima:03,machta:10a} might be another interesting contender in this
  respect \cite {wang:15,barash:16}.}\BibitemShut {Stop}%
\bibitem [{\citenamefont {Landau}\ and\ \citenamefont
  {Binder}(2015)}]{binder:book2}%
  \BibitemOpen
  \bibfield  {author} {\bibinfo {author} {\bibfnamefont {D.~P.}\ \bibnamefont
  {Landau}}\ and\ \bibinfo {author} {\bibfnamefont {K.}~\bibnamefont
  {Binder}},\ }\href@noop {} {\emph {\bibinfo {title} {A Guide to Monte Carlo
  Simulations in Statistical Physics}}},\ \bibinfo {edition} {4th}\ ed.\
  (\bibinfo  {publisher} {Cambridge University Press},\ \bibinfo {address}
  {Cambridge},\ \bibinfo {year} {2015})\BibitemShut {NoStop}%
\bibitem [{\citenamefont {Bittner}\ \emph {et~al.}(2008)\citenamefont
  {Bittner}, \citenamefont {Nussbaumer},\ and\ \citenamefont
  {Janke}}]{bittner:08}%
  \BibitemOpen
  \bibfield  {author} {\bibinfo {author} {\bibfnamefont {E.}~\bibnamefont
  {Bittner}}, \bibinfo {author} {\bibfnamefont {A.}~\bibnamefont {Nussbaumer}},
  \ and\ \bibinfo {author} {\bibfnamefont {W.}~\bibnamefont {Janke}},\ }\href
  {http://dx.doi.org/10.1103/physrevlett.101.130603} {\bibfield  {journal}
  {\bibinfo  {journal} {Phys. Rev. Lett.}\ }\textbf {\bibinfo {volume} {101}},\
  \bibinfo {pages} {130603} (\bibinfo {year} {2008})}\BibitemShut {NoStop}%
\bibitem [{\citenamefont {Katzgraber}\ \emph {et~al.}(2006)\citenamefont
  {Katzgraber}, \citenamefont {Trebst}, \citenamefont {Huse},\ and\
  \citenamefont {Troyer}}]{katzgraber:06}%
  \BibitemOpen
  \bibfield  {author} {\bibinfo {author} {\bibfnamefont {H.~G.}\ \bibnamefont
  {Katzgraber}}, \bibinfo {author} {\bibfnamefont {S.}~\bibnamefont {Trebst}},
  \bibinfo {author} {\bibfnamefont {D.~A.}\ \bibnamefont {Huse}}, \ and\
  \bibinfo {author} {\bibfnamefont {M.}~\bibnamefont {Troyer}},\ }\href
  {http://dx.doi.org/10.1088/1742-5468/2006/03/p03018} {\bibfield  {journal}
  {\bibinfo  {journal} {J. Stat. Mech.: Theory and Exp.}\ }\textbf {\bibinfo
  {volume} {2006}},\ \bibinfo {pages} {P03018} (\bibinfo {year}
  {2006})}\BibitemShut {NoStop}%
\bibitem [{\citenamefont {Kofke}(2002)}]{kofk02}%
  \BibitemOpen
  \bibfield  {author} {\bibinfo {author} {\bibfnamefont {D.~A.}\ \bibnamefont
  {Kofke}},\ }\href@noop {} {\bibfield  {journal} {\bibinfo  {journal} {J.
  Chem. Phys.}\ }\textbf {\bibinfo {volume} {117}},\ \bibinfo {pages} {6911}
  (\bibinfo {year} {2002})}\BibitemShut {NoStop}%
\bibitem [{\citenamefont {Kone}\ and\ \citenamefont {Kofke}(2005)}]{kon05}%
  \BibitemOpen
  \bibfield  {author} {\bibinfo {author} {\bibfnamefont {A.}~\bibnamefont
  {Kone}}\ and\ \bibinfo {author} {\bibfnamefont {D.~A.}\ \bibnamefont
  {Kofke}},\ }\href@noop {} {\bibfield  {journal} {\bibinfo  {journal} {J.
  Chem. Phys.}\ }\textbf {\bibinfo {volume} {122}},\ \bibinfo {pages} {206101}
  (\bibinfo {year} {2005})}\BibitemShut {NoStop}%
\bibitem [{\citenamefont {Predescu}\ \emph {et~al.}(2004)\citenamefont
  {Predescu}, \citenamefont {Predescu},\ and\ \citenamefont
  {Ciobanu}}]{pred04}%
  \BibitemOpen
  \bibfield  {author} {\bibinfo {author} {\bibfnamefont {C.}~\bibnamefont
  {Predescu}}, \bibinfo {author} {\bibfnamefont {M.}~\bibnamefont {Predescu}},
  \ and\ \bibinfo {author} {\bibfnamefont {C.~V.}\ \bibnamefont {Ciobanu}},\
  }\href@noop {} {\bibfield  {journal} {\bibinfo  {journal} {J. Chem. Phys.}\
  }\textbf {\bibinfo {volume} {120}},\ \bibinfo {pages} {4119} (\bibinfo {year}
  {2004})}\BibitemShut {NoStop}%
\bibitem [{\citenamefont {Sabo}\ \emph {et~al.}(2008)\citenamefont {Sabo},
  \citenamefont {Meuwly}, \citenamefont {Freeman},\ and\ \citenamefont
  {Doll}}]{sabo08}%
  \BibitemOpen
  \bibfield  {author} {\bibinfo {author} {\bibfnamefont {D.}~\bibnamefont
  {Sabo}}, \bibinfo {author} {\bibfnamefont {M.}~\bibnamefont {Meuwly}},
  \bibinfo {author} {\bibfnamefont {D.~L.}\ \bibnamefont {Freeman}}, \ and\
  \bibinfo {author} {\bibfnamefont {J.}~\bibnamefont {Doll}},\ }\href@noop {}
  {\bibfield  {journal} {\bibinfo  {journal} {J. Chem. Phys.}\ }\textbf
  {\bibinfo {volume} {128}},\ \bibinfo {pages} {174109} (\bibinfo {year}
  {2008})}\BibitemShut {NoStop}%
\bibitem [{\citenamefont {Neirotti}\ \emph {et~al.}(2000)\citenamefont
  {Neirotti}, \citenamefont {Calvo}, \citenamefont {Freeman},\ and\
  \citenamefont {Doll}}]{neiro00}%
  \BibitemOpen
  \bibfield  {author} {\bibinfo {author} {\bibfnamefont {J.}~\bibnamefont
  {Neirotti}}, \bibinfo {author} {\bibfnamefont {F.}~\bibnamefont {Calvo}},
  \bibinfo {author} {\bibfnamefont {D.~L.}\ \bibnamefont {Freeman}}, \ and\
  \bibinfo {author} {\bibfnamefont {J.}~\bibnamefont {Doll}},\ }\href@noop {}
  {\bibfield  {journal} {\bibinfo  {journal} {J. Chem. Phys.}\ }\textbf
  {\bibinfo {volume} {112}},\ \bibinfo {pages} {10340} (\bibinfo {year}
  {2000})}\BibitemShut {NoStop}%
\bibitem [{\citenamefont {Calvo}(2005)}]{calvo05}%
  \BibitemOpen
  \bibfield  {author} {\bibinfo {author} {\bibfnamefont {F.}~\bibnamefont
  {Calvo}},\ }\href@noop {} {\bibfield  {journal} {\bibinfo  {journal} {J.
  Chem. Phys.}\ }\textbf {\bibinfo {volume} {123}},\ \bibinfo {pages} {124106}
  (\bibinfo {year} {2005})}\BibitemShut {NoStop}%
\bibitem [{\citenamefont {Brenner}\ \emph {et~al.}(2007)\citenamefont
  {Brenner}, \citenamefont {Sweet}, \citenamefont {VonHandorf},\ and\
  \citenamefont {Izaguirre}}]{brenner07}%
  \BibitemOpen
  \bibfield  {author} {\bibinfo {author} {\bibfnamefont {P.}~\bibnamefont
  {Brenner}}, \bibinfo {author} {\bibfnamefont {C.~R.}\ \bibnamefont {Sweet}},
  \bibinfo {author} {\bibfnamefont {D.}~\bibnamefont {VonHandorf}}, \ and\
  \bibinfo {author} {\bibfnamefont {J.~A.}\ \bibnamefont {Izaguirre}},\
  }\href@noop {} {\bibfield  {journal} {\bibinfo  {journal} {J. Chem. Phys.}\
  }\textbf {\bibinfo {volume} {126}},\ \bibinfo {pages} {074103} (\bibinfo
  {year} {2007})}\BibitemShut {NoStop}%
\bibitem [{\citenamefont {Katzgraber}(2009)}]{katzgraber:09a}%
  \BibitemOpen
  \bibfield  {author} {\bibinfo {author} {\bibfnamefont {H.~G.}\ \bibnamefont
  {Katzgraber}},\ }\href {http://arxiv.org/abs/0905.1629} {\enquote {\bibinfo
  {title} {Introduction to {M}onte {C}arlo methods},}}
preprint \Eprint {http://arxiv.org/abs/0905.1629} {arXiv:0905.1629}
   (\bibinfo {year}
  {2009})\BibitemShut {NoStop}%
\bibitem [{Note2()}]{Note2}%
  \BibitemOpen
  \bibinfo {note} {In a somewhat simpler setup, one could also just monitor the
  swap acceptance rates.}\BibitemShut {Stop}%
\bibitem [{Note3()}]{Note3}%
  \BibitemOpen
  \bibinfo {note} {Note that we did not fully follow the optimization procedure
  described above, but only considered a few discrete choices of $\eta
  $.}\BibitemShut {Stop}%
\bibitem [{\citenamefont {Khoshbakht}\ and\ \citenamefont
  {Weigel}(2017)}]{khoshbakht:17a}%
  \BibitemOpen
  \bibfield  {author} {\bibinfo {author} {\bibfnamefont {H.}~\bibnamefont
  {Khoshbakht}}\ and\ \bibinfo {author} {\bibfnamefont {M.}~\bibnamefont
  {Weigel}},\ }\href {http://arxiv.org/abs/1710.01670} {\enquote {\bibinfo
  {title} {Domain-wall excitations in the two-dimensional {I}sing spin
  glass},}}\ \Eprint
{http://arxiv.org/abs/1710.01670} {arXiv:1710.01670}  (\bibinfo {year} {2017})
\BibitemShut {NoStop}%
\bibitem [{Note4()}]{Note4}%
  \BibitemOpen
  \bibinfo {note} {Note that Eq.~(\ref {eq:success}) is actually valid at the
  level of individual disorder samples, but it also typically provides a good
  approximation if used at the level of disorder averages \cite
  {weigel:06b}.}\BibitemShut {Stop}%
\bibitem [{\citenamefont {Weigel}(2018)}]{weigel:18}%
  \BibitemOpen
  \bibfield  {author} {\bibinfo {author} {\bibfnamefont {M.}~\bibnamefont
  {Weigel}},\ }\enquote {\bibinfo {title} {{M}onte {C}arlo methods for
  massively parallel computers},}\ in\ \href@noop {} {\emph {\bibinfo
  {booktitle} {Order, Disorder and Criticality}}},\ Vol.~\bibinfo {volume}
  {5},\ \bibinfo {editor} {edited by\ \bibinfo {editor} {\bibfnamefont
  {Y.}~\bibnamefont {Holovatch}}}\ (\bibinfo  {publisher} {World Scientific},\
  \bibinfo {address} {Singapore},\ \bibinfo {year} {2018})\ pp.\ \bibinfo
  {pages} {271--340}\BibitemShut {NoStop}%
\bibitem [{\citenamefont {Kumar}\ \emph {et~al.}()\citenamefont {Kumar},
  \citenamefont {Gross}, \citenamefont {Weigel},\ and\ \citenamefont
  {Janke}}]{gross:18}%
  \BibitemOpen
  \bibfield  {author} {\bibinfo {author} {\bibfnamefont {R.}~\bibnamefont
  {Kumar}}, \bibinfo {author} {\bibfnamefont {J.}~\bibnamefont {Gross}},
  \bibinfo {author} {\bibfnamefont {M.}~\bibnamefont {Weigel}}, \ and\ \bibinfo
  {author} {\bibfnamefont {W.}~\bibnamefont {Janke}},\ }\href@noop {} {\enquote
  {\bibinfo {title} {Effective {GPU} implementation of parallel tempering for
  spin glasses and random-field models},}\ }\bibinfo {note} {in
  preparation}\BibitemShut {NoStop}%
\bibitem [{\citenamefont {Binder}(1983{\natexlab{b}})}]{binder:83a}%
  \BibitemOpen
  \bibfield  {author} {\bibinfo {author} {\bibfnamefont {K.}~\bibnamefont
  {Binder}},\ }in\ \href@noop {} {\emph {\bibinfo {booktitle} {Phase
  Transitions and Critical Phenomena}}},\ Vol.~\bibinfo {volume} {8},\ \bibinfo
  {editor} {edited by\ \bibinfo {editor} {\bibfnamefont {C.}~\bibnamefont
  {Domb}}\ and\ \bibinfo {editor} {\bibfnamefont {J.~L.}\ \bibnamefont
  {Lebowitz}}}\ (\bibinfo  {publisher} {Academic Press},\ \bibinfo {address}
  {New York},\ \bibinfo {year} {1983})\ pp.\ \bibinfo {pages}
  {1--144}\BibitemShut {NoStop}%
\bibitem [{\citenamefont {Sepp{\"{a}}l{\"{a}}}\ \emph
  {et~al.}(1998)\citenamefont {Sepp{\"{a}}l{\"{a}}}, \citenamefont
  {Pet{\"{a}}j{\"{a}}},\ and\ \citenamefont {Alava}}]{seppala:98}%
  \BibitemOpen
  \bibfield  {author} {\bibinfo {author} {\bibfnamefont {E.~T.}\ \bibnamefont
  {Sepp{\"{a}}l{\"{a}}}}, \bibinfo {author} {\bibfnamefont {V.}~\bibnamefont
  {Pet{\"{a}}j{\"{a}}}}, \ and\ \bibinfo {author} {\bibfnamefont {M.~J.}\
  \bibnamefont {Alava}},\ }\href {http://dx.doi.org/10.1103/PhysRevE.58.R5217}
  {\bibfield  {journal} {\bibinfo  {journal} {Phys. Rev. E}\ }\textbf {\bibinfo
  {volume} {58}},\ \bibinfo {pages} {R5217} (\bibinfo {year}
  {1998})}\BibitemShut {NoStop}%
\bibitem [{\citenamefont {Hukushima}\ and\ \citenamefont
  {Iba}(2003)}]{hukushima:03}%
  \BibitemOpen
  \bibfield  {author} {\bibinfo {author} {\bibfnamefont {K.}~\bibnamefont
  {Hukushima}}\ and\ \bibinfo {author} {\bibfnamefont {Y.}~\bibnamefont
  {Iba}},\ }\href {http://dx.doi.org/10.1063/1.1632130} {\bibfield  {journal}
  {\bibinfo  {journal} {AIP Conf. Proc.}\ }\textbf {\bibinfo {volume} {690}},\
  \bibinfo {pages} {200} (\bibinfo {year} {2003})}\BibitemShut {NoStop}%
\bibitem [{\citenamefont {Machta}(2010)}]{machta:10a}%
  \BibitemOpen
  \bibfield  {author} {\bibinfo {author} {\bibfnamefont {J.}~\bibnamefont
  {Machta}},\ }\href {http://link.aps.org/doi/10.1103/PhysRevE.82.026704}
  {\bibfield  {journal} {\bibinfo  {journal} {Phys. Rev. E}\ }\textbf {\bibinfo
  {volume} {82}},\ \bibinfo {pages} {026704} (\bibinfo {year}
  {2010})}\BibitemShut {NoStop}%
\bibitem [{\citenamefont {Barash}\ \emph {et~al.}(2017)\citenamefont {Barash},
  \citenamefont {Weigel}, \citenamefont {Borovsk\'{y}}, \citenamefont {Janke},\
  and\ \citenamefont {Shchur}}]{barash:16}%
  \BibitemOpen
  \bibfield  {author} {\bibinfo {author} {\bibfnamefont {L.~Y.}\ \bibnamefont
  {Barash}}, \bibinfo {author} {\bibfnamefont {M.}~\bibnamefont {Weigel}},
  \bibinfo {author} {\bibfnamefont {M.}~\bibnamefont {Borovsk\'{y}}}, \bibinfo
  {author} {\bibfnamefont {W.}~\bibnamefont {Janke}}, \ and\ \bibinfo {author}
  {\bibfnamefont {L.~N.}\ \bibnamefont {Shchur}},\ }\href@noop {} {\bibfield
  {journal} {\bibinfo  {journal} {Comput. Phys. Commun.}\ }\textbf {\bibinfo
  {volume} {220}},\ \bibinfo {pages} {341} (\bibinfo {year}
  {2017})}\BibitemShut {NoStop}%
\end{thebibliography}

%

\end{document}